\documentclass[12pt, aps, prd, reprint, twocolumn, sort&compress, superscriptaddress, showpacs, nofootinbib, preprintnumbers]{revtex4-1}

\usepackage{graphicx}
\usepackage{dcolumn}
\usepackage{bm}
\usepackage{amsmath}
\usepackage{amssymb}
\usepackage{dsfont}
\usepackage{amsfonts}
\usepackage[ngerman,UKenglish]{babel}
\usepackage{hyperref}
\usepackage{nicefrac}
\usepackage{verbatim}
\usepackage{bbm}
\usepackage{color}

\def\dbar{{\mathchar'26\mkern-11mu  d}}

\newcommand{\bx}{\boldsymbol{x}}
\newcommand{\by}{\boldsymbol{y}}
\newcommand{\bz}{\boldsymbol{z}}

\newcommand{\bk}{\boldsymbol{k}}
\newcommand{\bp}{\boldsymbol{p}}
\newcommand{\bq}{\boldsymbol{q}}

\newcommand{\vx}{\boldsymbol{x}}
\newcommand{\vB}{\boldsymbol{B}}
\newcommand{\vD}{\boldsymbol{D}}
\newcommand{\cD}{\cal{D}}
\newcommand{\cN}{\cal{N}}

\newcommand{\vk}{\boldsymbol{k}}
\newcommand{\vz}{\boldsymbol{z}}
\newcommand{\vy}{\boldsymbol{y}}
\newcommand{\vK}{\boldsymbol{K}}
\newcommand{\vA}{\boldsymbol{A}}

\newcommand{\vp}{\boldsymbol{p}}
\newcommand{\vq}{\boldsymbol{q}}

\newcommand{\il}{\int\limits}

\newcommand{\Id}{ \mathbbm{1} }
\newcommand{\lk}{\left(}
\newcommand{\rk}{\right)}
\renewcommand{\vec}[1]{\mbox{\boldmath$#1$\unboldmath}}

\begin{document}

\title{Quark Sector of the QCD Groundstate in Coulomb Gauge}

\author{M.~Pak}
\affiliation{Institut f\"ur Physik, FB Theoretische Physik, Universit\"at Graz, Universit\"atsplatz 5,
8010 Graz, Austria}
\email{markus.pak@uni-graz.at}
\author{H.~Reinhardt}
\affiliation{Institut f\"ur Theoretische Physik, Universit\"at T\"ubingen, Auf der Morgenstelle 14,
72076 T\"ubingen, Germany}
\email{hugo.reinhardt@uni-tuebingen.de}

\begin{abstract}
The variational approach to Yang-Mills theory in Coulomb gauge is extended to full
QCD. For the quark sector we use a trial wave functional, which goes beyond the previously used BCS-type state and which
  explicitly contains the
coupling of the quarks to transverse gluons.
This quark wave functional contains two variational kernels: One is related to the quark condensate and occurs already in the BCS-type states. The other represents the
form factor of the coupling of the quarks to the transverse gluons. Minimization of the energy density with respect to
these kernels results in two coupled integral (gap) equations. These equations are solved numerically using
the confining part of the non-Abelian color Coulomb potential and
the lattice static gluon propagator as input.
With the additional coupling of quarks to
transverse gluons included the low energy chiral properties increase
substantially towards their phenomenological values. We obtain a
reasonable description of the chiral condensate, which for a vanishing current quark mass is obtained in the range of 190 --
235 MeV. The coupling of the quarks to the transverse gluons enhances the
constituent quark mass
by about 60\% in comparison to the pure BCS ansatz.

\end{abstract}

\maketitle

\section{Introduction}

Understanding the low energy sector of QCD is one of the major challenges of particle physics. This sector is characterized by
two non-perturbative phenomena: confinement and chiral symmetry breaking. 
Color confinement is assumed to be essentially due to the
gluon sector. In recent years substantial progress in understanding the
low energy sector of Yang-Mills theory has been achieved within non-perturbative continuum approaches. Among these is a
variational approach to Yang-Mills theory in Coulomb gauge, Ref.~\cite{Feuchter:2004mk}. There is a long history of the variational treatment of the
Yang-Mills vacuum sector in Coulomb gauge, see for example Refs.~\cite{Schutte:1985sd,Szczepaniak:2001rg}. Our approach differs from previous work in the choice of
the trial wave functional and, more importantly, in the full inclusion of the Faddeev-Popov determinant and in the renormalization procedure, see Refs.~\cite{Greensite:2011pj} 
for more details. Our variational approach has given a quite decent description of the infrared sector of Yang-Mills theory as,
for example, a linearly rising non-Abelian Coulomb potential \cite{Epple:2006hv}, an infrared diverging gluon energy (expressing confinement)
\cite{Feuchter:2004mk,Epple:2006hv} in accord with
lattice data \cite{Burgio:2008jr}, an infrared finite running coupling constant
\cite{Schleifenbaum:2006bq},
a perimeter law for the 't Hooft loop \cite{Reinhardt:2007wh},
an area law for the Wilson loop \cite{Pak:2009em} and a dielectric function of the Yang-Mills vacuum in
accord with the bag model picture \cite{Reinhardt:2008ek}.
The obtained infrared behavior of ghost
and gluon propagators were also found in a functional renormalization group approach \cite{Leder:2010ji}
and supported by lattice calculation \cite{Burgio:2008jr, Burgio:2012bk}. Furthermore, recently
the variational approach of Ref.~\cite{Feuchter:2004mk}
was extended to finite temperatures \cite{Reinhardt:2011hq}, \cite{Heffner:2012sx}. A critical temperature in the range of
 $T_{\textsc{C}} = 270 - 290 \,$ MeV was obtained \cite{Heffner:2012sx},
which is in the range of the lattice data \cite{Fingberg:1992ju, Kaczmarek:2002mc, Lucini:2005vg}. 
A similar transition temperature of $T_{\textsc{C}} \approx 270 \,$ MeV
was also found from the effective potential of
the Polyakov loop \cite{Reinhardt:2012qe, Reinhardt:2013iia}.

In the present paper we extend the variational approach in Coulomb gauge to full QCD. The low energy quark sector of QCD is dominated by
chiral symmetry and its spontaneous breaking. For $N_f$ massless quark flavors QCD is invariant under separate global flavor rotations of
the left- and right-handed quarks. In the vacuum the $U_{\textsc{L}} (N_f) \times U_{\textsc{R}} (N_f)$ symmetry group is spontaneously broken to the diagonal
vector group $U_{\textsc{V}} (N_f)$ by quark condensation $\langle \overline{q} q \rangle \neq 0$, resulting in the appearance of $N^2_f$ pseudoscalar
massless Goldstone bosons corresponding to the generators of the coset $U_{\textsc{A}} (N_f) = U_{\textsc{L}} (N_f) \times U_{\textsc{R}} (N_f) / U_{\textsc{V}} (N_f)$. Chiral
symmetry is a good starting point for $N_f = 3$ quark flavors $u, d, s$. When the small current quark masses
 of the light flavors are included
chiral symmetry is explicitly broken and the (would-be) Goldstone bosons aquire a finite mass. Finally, the chiral anomaly breaks the $U_{\textsc{A}}(N_f)$
down to $S U_{\textsc{A}}(N_f)$ thereby providing an extra mass to the Goldstone boson of the $U_{\textsc{A}}(1)$ generator, which
corresponds to the $\eta'$-meson.

The mechanism of spontaneous breaking of chiral symmetry was first investigated in effective models of the Nambu--Jona-Lasinio type,
Refs.~\cite{Nambu:1961tp, Nambu:1961fr}, which could explain the quark condensation in analogy to the emergence of
Cooper pairs in
superconductors. The Nambu--Jona-Lasinio model has been successful not only in explaining the mechanism of spontaneous breaking of
chiral symmetry but also in describing the low-energy data of the light pseudoscalar mesons. For this purpose the Nambu--Jona-Lasinio model
was bosonized and the resulting effective meson theory was worked out in a gradient expansion \cite{Ebert:1985kz}.
Inspired by these model studies, the quark sector of QCD was treated in the variational approach in Coulomb
gauge assuming BCS-type trial quark
wave functionals and a purely confining static quark potential, Refs.~\cite{Finger:1981gm,LeYaouanc:1983iy,Adler:1984ri}.
 In these calculations the
coupling of the transverse gluons to the quarks was neglected resulting in substantially 
too small values of the quark condensate,
 the constituent
mass and pion decay constant. To improve these results in Refs.~\cite{Alkofer:1988tc,LlanesEstrada:2004wr} an additional four quark interaction mediated by static transverse
gluons was introduced, see also Ref.~\cite{Fontoura:2012mz}. Of course, an additional attractive interaction will enhance the amount of chiral symmetry breaking.
However, such a four-fermion interaction mediated by transverse gluons is not in the QCD Hamiltonian in Coulomb gauge in the first place.
Rather does this Hamiltonian contain an explicit coupling of the quarks to the transverse gluons and it is a priori not clear, to which extend
this coupling can be simulated by a static four-quark interaction. The quark-gluon coupling of the QCD Hamiltonian escapes the variational approach
when a BCS-type trial wave functional is used. In the present paper we go beyond the BCS type of approximations considered previously and use a
quark wave functional, which explicitly includes the coupling to the transverse gluons. The form factor of this coupling is treated as a
variational kernel determined from the minimization of the energy. First results obtained with this wave functional have been already reported in Ref.~\cite{Pak:2011wu}.
Here we give a more detailed and complete account of the variational approach to QCD in Coulomb gauge with the trial wave functional
proposed in Ref.~\cite{Pak:2011wu}.   

Though, this approach can, in principle, be carried out in a fully self-consistent manner, minimizing simultaneously the energy with respect to all
kernels of the trial wave functional, in order to keep the formal exposition sufficiently transparent in the present paper we will focus on
the quark sector and do mainly a quenched calculation, ignoring the back-reaction of the quarks on the gluon sector, i.e. we will use the results
obtained in the variational approach to Yang-Mills theory, in particular the gluon dispersion relation, as an input.
However, in Sect.~\ref{XIunqu}
 we will study the effect of the quarks on the gluon propagator.

The organization of the paper is as follows: In Section \ref{Hamiltonian-sec} we review the Hamiltonian
approach to QCD in Coulomb gauge. In Section \ref{Yang-Mills-sec} we
briefly collect results gained in the pure Yang-Mills sector of
QCD, which are needed as input for the present work. In Section \ref{Gen-functional-sec} we
present our quark vacuum wave functional, which includes the
interaction of quarks with transverse gluon fields and which was originally proposed in Ref.~\cite{Pak:2011wu}.
The Dirac and color structure
of the variational kernels are specified.
In Section \ref{Gen-functional-secII} we set up the QCD
generating functional in order to compute the various $n$-point
functions of the theory. In Section \ref{QuarkProp} the quark
propagator and related chiral quantities are expressed in terms of the variational functions.
In Sections \ref{Dirac-E} we start our variational analysis by computing the
energy density of the quarks and carry out the variation of the latter with respect to the two kernels of the wave
functional, resulting in two coupled gap equations. These equations are studied in Sect.~\ref{Asympt-sect} in the IR- and UV-regime.
In a first variational analysis we demonstrate in Sect.~\ref{sectionIX} that within
the present approach the coupling of the quarks to the transverse gluons alone cannot induce spontaneous
breaking of chiral symmetry. The full variational calculation with the color Coulomb potential included is carried
out in Sect.~\ref{Xnumres}. Here we solve the corresponding coupled gap equation numerically and calculate the
chiral properties of the quarks. In Sect.~\ref{XIunqu} we give an estimate of the unquenching effects on the gluon propagator.
 In the last section \ref{XIIsum} we
summarize our findings, present  our conclusions and give an outlook on future studies.

\section{Hamiltonian approach to QCD in Coulomb gauge}
\label{Hamiltonian-sec}
The Hamiltonian approach to QCD is based on the canonical quantization in Weyl gauge $A_0 = 0$, which leaves the spatial components
of the gauge field $\vA (\vx)$ as independent coordinates and results in the Hamiltonian
\begin{align}
\label{152-x1}
H_{\textsc{QCD}} = H_{\textsc{YM}} + H_{\textsc{F}} \, ,
\end{align}
where
\begin{align}
\label{157-x2}
H_{\textsc{YM}} = \frac{1}{2} \int d^3 x \lk \vec{\Pi}^2 (\vx) + \vB^2(\vx) \rk
\end{align}
is the Yang-Mills Hamiltonian with $\vB(\vx)$ being the non-Abelian magnetic field and
\begin{align}
\label{162-x3}
\vec{\Pi}^a (\bx) = \frac{\delta}{i \delta \vA^a (\vx)}
\end{align}
being the conjugate momentum operator. Furthermore
\begin{align}
\label{167-x4}
H_{\textsc{F}} = \int d^3 x \, \psi^\dagger (\bx) \lk -i  \vec{\alpha} \cdot \vD + \beta m_0 \rk \psi (\vx)
\end{align}
is the Hamilton operator of the quark field $\psi (\vx)$, which satisfies the anti-commutation relation
\begin{align}
\label{172}
\left\{ \psi (\vx) \, , \, \psi^\dagger (\vy) \right\} = \delta (\vx - \vy) \, .
\end{align}
Here $\vec{\alpha}, \beta$ are the usual Dirac matrices satisfying
$\{\alpha_i,\alpha_j\} = \delta_{i j}$ and $\{\beta, \alpha_i \} = 0$, $m_0$ is the current quark mass and
\begin{align}
\label{177-x5}
\vD = \vec{\partial} - i g T^a \vA^a
\end{align}
is the covariant derivative with $T^a$ being the (Hermitian) generators of the gauge group $SU (N)$ in the fundamental
representation. We have
suppressed here the Lorentz,
color and flavor indices of the quarks.
For the present consideration it is sufficient to consider a single
flavor so that flavor becomes irrelevant.
Due to the use of Weyl gauge Gauss' law escapes the equaton of motion and has to be imposed as constraint
to the wave functional $\phi [\vA, \psi]$ 
\begin{align}
\label{186-x6}
\lk \hat{\vD}^{a b} \vec{\Pi}^b \rk (\vx) \phi [\vA, \psi] = \rho^a_{\textsc{F}} (\vx) \phi [\vA, \psi] \, .
\end{align}
Here
\begin{align}
\label{191-x7}
\hat{\vD}^{ab} = \delta^{ab} + g f^{acb} \vA^c
\end{align}
is the covariant derivative in the adjoint representation and
\begin{align}
\label{196-x8}
\rho^a_{\textsc{F}} (\vx) = \psi^\dagger (\vx) T^a \psi (\vx)
\end{align}
are the color charge densities of the quarks. The operator $\hat{\vD}^{a b} \vec{\Pi}^b$ in Gauss' law is the generator of time-independent gauge
transformations, which are not fixed by Weyl gauge. We fix this residual gauge freedom by choosing the Coulomb gauge
\begin{align}
\label{202}
\vec{\partial} \vA = 0 \, .
\end{align}
In this gauge Gauss' law can be explicitly resolved, which results in the gauge fixed Hamiltonian \cite{Christ:1980ku}
\begin{align}
\label{207-x9}
H^C_{\textsc{QCD}} = H^C_{\textsc{YM}} + H_{\textsc{F}} + H_{\textsc{C}} \, ,
\end{align}
where
$H_{\textsc{F}}$ is the same as in (\ref{167-x4}) except that the gauge field is now transversal:
\begin{align}
\label{A-perp}
 A_i^{\perp a}(\bx) = t_{i j}(\bx) A_j^{a}(\bx) \; ,  
 \end{align}
 with
 \begin{align}
 t_{i j}(\bx) = \int  \frac{d^3 p}{(2\pi)^3} \left(\delta_{i j} -\hat{p}_i \hat{p}_j \right) e^{i \bp \cdot \bx} \; , \quad \hat{p}_i=\frac{p_i}{|\bp|} \; . 
 \end{align}
Furthermore,
\begin{widetext}
\begin{align}
\label{212-x10}
H^C_{\textsc{YM}} = \frac{1}{2} \int d^3 x \lk J^{- 1}[\vA^{\perp}] \Pi^{\perp a}_i (\vx) J[\vA^{\perp}] \Pi^{\perp a}_i(\vx) + B_i^a(\bx)^2 \rk 
\end{align}
\end{widetext}
is the Hamiltonian of the transverse gluons, where
\begin{align}
\label{264-xx}
\Pi_i^{\perp a}(\bx) = \frac{\delta}{i \delta A_i^{\perp a}(\bx)} = t_{i j}(\bx) \Pi_j^a(\bx) ,
\end{align}
and
\begin{align}
\label{Ch-251-13}
J [\vA^{\perp}] = \mbox{Det}(- \boldsymbol{\hat{D}} \boldsymbol{\partial})
\end{align}
 is the Faddeev-Popov determinant.
 The Coulomb term
\begin{align}
\label{232}
H_{\textsc{C}} \! = \!\frac{g^2}{2} \! \int d^3 x \int d^3 y J^{- 1}[\vA^{\perp}] \rho^a (\vx) F^{ab} (\vx, \vy) J[\vA^{\perp}] \rho^b (\vy)
\end{align}
arises from the kinetic energy of the longitudinal modes after resolving Gauss' law. Here
\begin{align}
\label{237}
F^{a b}(\bx, \by) = \langle \vx a | (- \boldsymbol{\hat{D}} \boldsymbol{\partial})^{- 1} (- \boldsymbol{\partial}^2) (- \boldsymbol{\hat{D}} \boldsymbol{\partial})^{- 1} | \vy b\rangle
\end{align}
is the so-called Coulomb  kernel and
\begin{align}
\label{242}
\rho^a(\vx) = \rho^a_{\textsc{YM}} (\vx) + \rho^a_{\textsc{F}} (\vx)
\end{align}
is the total color charge density, which contains besides the charge of the quarks, $\rho^a_{\textsc{F}} (\bx)$, Eq.~(\ref{196-x8}), also the color charge
of the gauge field
\begin{align}
\label{248}
\rho^a_{\textsc{YM}} (\vx) = - f^{abc} \vA^{\perp b}(\bx) \boldsymbol{\Pi}^{\perp c}(\bx) \, .
\end{align}
In the rest of the paper 
we work exclusively in Coulomb gauge and from now on we will  omit the transversality sign 
attached to the gauge field.

We are interested here in the ground state wave functional of QCD, which we will determine from a variational calculation.
Without loss of generality we can choose the trial wave functional in the coordinate representation of the gauge field in the form
\begin{align}
\label{270-y1}
\langle \vA | \phi \rangle = \phi_{\textsc{YM}} (\vA) | \phi_{\textsc{F}} (\vA) \rangle \, .
\end{align}
Here $| \phi_{\textsc{F}} (\vA) \rangle$ is the wave functional of the Dirac vacuum of the quarks in the presence of the gauge field and $\phi_{\textsc{YM}} (\vA)$ is the
wave functional of the Yang-Mills sector. We have chosen here the coordinate representation for the Yang-Mills part of the wave functional
$\phi_{\textsc{YM}}(\vA)$, while the fermion wave functional $|\phi_{\textsc{F}}(\vA) \rangle$ is chosen as ket-vector in
Fock space. Note $|\phi_{\textsc{F}} (\vA) \rangle$,
depending on the gauge field, contains the  full coupling of the quarks to gluons. 

The expectation value of an observable $O [\vA, \psi]$ in the state (\ref{270-y1}) is given by
\begin{widetext}
\begin{align}
\label{280-y2}
\langle O [\vA, \psi] \rangle = \int {\cD} \vA \, J (A) \, \phi^*_{\textsc{YM}}(\vA) \left\langle \phi_F (\vA) | O [\vA, \psi] | \phi_{\textsc{F}} (\vA) \right\rangle \, \phi_{\textsc{YM}} (\vA) \, .
\end{align}
\end{widetext}
Note the presence of the Faddeev-Popov determinant $J [A]$, Eq.~(\ref{Ch-251-13}), in the integration measure.
 In principle, the fermion wave functional $| \phi_F (\vA) \rangle$ could also be
expressed in a ``coordinate'' representation, i.e. in terms of Grassmann variables.
Then the scalar product (\ref{280-y2}) would also contain the integration over Grassmann fields. In the
present case it is, however, more convenient to represent the fermionic wave functional in 
second quantized form as a vector in Fock space,
see Sect.~\ref{Gen-functional-sec}.

With the color charge density $\rho^a(\bx)$, Eq.~(\ref{242}),
 being a sum of a gluonic and a quark part, the Coulomb Hamiltonian $H_{\textsc{C}}$, Eq.~(\ref{232}),
 can be split up as
\begin{align}
\label{289-y3}
H_{\textsc{C}} = H^{\textsc{YM}}_{\textsc{C}} + H^{\textsc{coupl}}_{\textsc{C}} + H^{\textsc{F}}_{\textsc{C}} \, ,
\end{align}
where $H^{\textsc{YM}}_C$ and $H^{\textsc{F}}_C$ depend exclusively on the gauge field $\vA$ and the quark field $\psi$, respectively, while $H^{\textsc{coupl}}_C$ contains the coupling
between both fields. With this splitting we can write the full gauge fixed QCD  Hamiltonian (\ref{207-x9}) in the form
\begin{align}
H^C_{\textsc{QCD}} = \overline{H}^C_{\textsc{YM}} (\vA) + \overline{H}^C_\textsc{F} (\vA, \psi) \, ,
\end{align}
where
\begin{align}
\label{300}
\overline{H}^C_{\textsc{YM}} (\vA) = H^C_{\textsc{YM}} + H^{\textsc{YM}}_{\textsc{C}}
\end{align}
 contains exclusively the gauge field and is the Coulomb gauge fixed Hamiltonian of pure Yang-Mills theory (which was treated variationally
in Ref.~\cite{Feuchter:2004mk}),
while
\begin{align}
\label{H-quark-all}
\overline{H}^C_{\textsc{F}} (\vA, \psi) = H_{\textsc{F}} + H^{\textsc{F}}_{\textsc{C}} + H^{\textsc{coupl}}_{\textsc{C}}
\end{align}
 contains all terms, which depend on the quark field. In particular, it contains the coupling of the quarks to the gluons, see Eq.~(\ref{167-x4}).

In a full variational calculation one would minimize the full energy
\begin{align}
\label{301-y5}
\langle H^C_{\textsc{QCD}} \rangle \to \text{min.}
\end{align}
in the state (\ref{270-y1}). Here we do first a quenched calculation varying the fermionic part of the energy only
\begin{align}
\label{306-y6}
\langle \overline{H}^C_{\textsc{F}} \rangle \to \text{min.}
\end{align}
thereby keeping the Yang-Mills part $\phi_{\textsc{YM}} (\vA)$ of the wave functional (\ref{270-y1}) fixed to the
Yang-Mills vacuum state determined previously in Ref.~\cite{Epple:2006hv} from
\begin{align}
\label{312-y7}
\langle \phi_{\textsc{YM}} | \overline{H}^C_{\textsc{YM}} | \phi_{\textsc{YM}} \rangle \to \text{min.}
\end{align}
In the next section we will briefly summarize the essential results obtained within the variational approach to Yang-Mills theory (\ref{312-y7}), which we use as input
for the variational treatment of the fermionic sector, Eq.~(\ref{306-y6}).
\vspace{1cm}
\section{Variational results for the pure Yang-Mills sector of QCD}
\label{Yang-Mills-sec}
In Refs.~\cite{Feuchter:2004mk,Epple:2006hv} pure Yang-Mills theory has been treated in a variational approach in Coulomb gauge
using the following trial ansatz for the vacuum wave functional
\begin{widetext}
\begin{align}
 \label{YM-wave-functional}
\phi_{\textsc{YM}} (\vA) &= \langle \vA |\phi_{\textsc{YM}} \rangle \nonumber\\
& =
\frac{\mathcal{N}_{\textsc{G}}}{\sqrt{\mathcal{J}[\vA]}} \exp\left(
- \frac{1}{2} \int d^3 x \, \int d^3 y \, A^ a_i(\bx) \, t_{i j}(\bx) \omega(\bx,\by) \, A_j^a(\by)  \right) \; .
\end{align}
\end{widetext}
Here $\mathcal{J}[\vA]$ is the Faddeev-Popov determinant, Eq.~(\ref{Ch-251-13}), 
$\mathcal{N}_{\textsc{G}}$ is a
normalization factor fixed by requiring
$\langle \phi_{\textsc{YM}} |\phi_{\textsc{YM}} \rangle=1$ and $\omega(\bx,\by)$ is the variational
kernel. The advantage of this ansatz is that the Faddeev-Popov determinant $J [\vA]$, Eq.~(\ref{Ch-251-13}), drops out from the integration measure
 (\ref{280-y2}).
As a consequence
the (static) gluon propagator is just given by the inverse of the kernel $\omega$:
\begin{widetext}
\begin{align}
\label{gluon-prop}
D_{i j}^{a b}(\bx,\by) : = \langle A_i^a(\bx) A_j^b(\by) \rangle_{\textsc{G}} =
\delta^{ab} t_{ij} (\bx) D (\bx - \by) \, , \, D (\bx - \by) = \frac{1}{2} \omega^{- 1} (\vx, \vy)  \; ,
\end{align}
\end{widetext}
where
\begin{align}
\label{342}
\langle \ldots \rangle_{\textsc{G}} = \langle \phi_{\textsc{YM}} | \ldots | \phi_{\textsc{YM}} \rangle
\end{align}
denotes the expectation value in the pure Yang-Mills vacuum state $| \phi_{\textsc{YM}} \rangle$, Eq.~(\ref{YM-wave-functional}).

\begin{figure}
\originalTeX
\centerline{
\includegraphics[width=100mm]{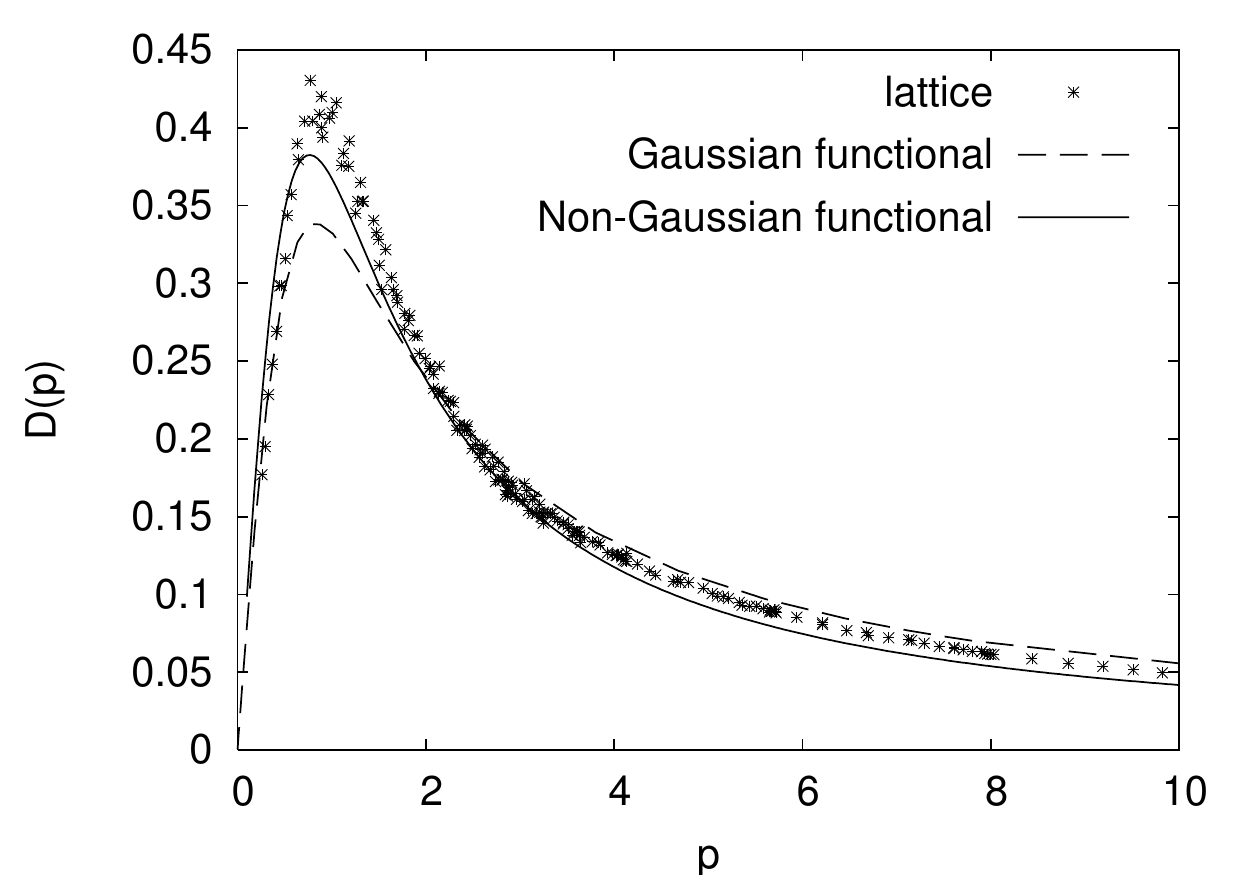}
}
\caption{\sl Gluon propagator $D(p)$. Data points are the lattice results
obtained in Ref.~\cite{Burgio:2008jr}. The dashed curve shows the results from the variational approach when a Gaussian vacuum is used. The full curve is the extension to
non-Gaussian wave functionals including up to quartic terms in the gauge field. The plot is from Ref.~\cite{Campagnari:2010wc}.}
\label{fig-Gribov}
\end{figure}

Variation of the pure gluonic energy density $\langle \overline{H}^C_{\textsc{YM}} \rangle_{\textsc{G}}$, Eq.~(\ref{300}), with
respect to the kernel $\omega$ yields a coupled system of integral
equations, see e.g.~Refs.~\cite{Feuchter:2004mk, Epple:2006hv}. These equations were solved analytically in the IR and UV asymptotic
momentum regions, Ref.~\cite{Schleifenbaum:2006bq}, as well as numerically in the
whole momentum regime, Refs.~\cite{Feuchter:2004mk, Epple:2006hv}. The gluon energy $\omega(\bp)$ is found to be IR divergent, expressing gluon confinement,
while it approaches for large momenta the photon energy, in accord with asymptotic freedom. Lattice calculations, Ref.~\cite{Burgio:2008jr}, confirm this
behavior and show that
 over the whole
 momentum range the gluon kernel $\omega(\bp)$ can be nicely fitted by Gribov's formula \cite{Gribov:1977wm}
 \begin{align}
 \label{Gribov-fit}
 \omega(\bp) = \sqrt{\bp^2 + \frac{M^4_{\textsc{G}}}{\bp^2}} \; ,
 \end{align}
 where $M_{\textsc{G}}$ is a mass scale referred to as Gribov mass. It was
 determined on the lattice in Ref.~\cite{Burgio:2008jr} and found to be given by
 \begin{align}
 \label{Gribov-Mass}
 M_{\textsc{G}} \approx 880 \, \text{MeV} = 2 \sqrt{\sigma_{\textsc{W}}} \; ,
 \end{align}
where $\sigma_{\textsc{W}}$ ($\sqrt{\sigma_{\textsc{W}}} = 440$ MeV) the Wilsonian string tension.
Fig.~\ref{fig-Gribov} shows the gluon propagator obtained in the variational approach together with lattice data. The results obtained with the
Gaussian wave functional (\ref{YM-wave-functional}) agree well with the lattice results in the IR and also in the UV but there are deviations in the mid-momentum
regime. These deviations substantially decrease when a non-Gaussian wave functional is used, which includes up to quartic terms in the
 exponent, see Ref.~\cite{Campagnari:2010wc}.

In later calculations we also need the vacuum expectation value of the Coulomb kernel $F (\vx, \vy)$, Eq.~(\ref{237}),
which represents the static potential between
(infinitely heavy) color point charges separated by a distance $r = |\vx - \vy|$
\begin{align}
\label{384}
g^2 \langle F (\vx, \vy) \rangle_{\textsc{G}} =: V_{\textsc{C}} (|\vx - \vy|) \, .
\end{align}
In the variational approach \cite{Epple:2006hv} one finds a potential which at large distances increases linearly
\begin{align}
\label{389}
V_{\textsc{C}} (r) = \sigma_{\textsc{C}} r \, , \quad \quad r \to \infty \, .
\end{align}
The same behavior is found on the lattice, Refs.~\cite{Voigt:2008rr,Nakagawa:2011ar,Greensite:2003xf}, with a Coulomb string tension $\sigma_{\textsc{C}}$ of
\begin{align}
\label{394}
\sigma_{\textsc{C}} \approx \left( 2 \ldots 3 \right) \sigma_{\textsc{W}} \, ,
\end{align}
where $\sigma_{\textsc{W}}$ is the Wilsonian string tension. In our approach the Coulomb string tension $\sigma_{\textsc{C}}$ is used to
 fix the scale. When we use the gluon propagator (\ref{Gribov-fit}) as input there is a second dimensionful input quantity:
the Gribov mass $M_G$.
These two
quantities are, however, not independent of each other. In the approximation
\begin{widetext}
\begin{align}
\label{num-1234}
\left\langle \lk - \boldsymbol{\hat{D}} \boldsymbol{\partial} \rk^{- 1} \lk - \boldsymbol{\partial}^2 \rk \lk - \boldsymbol{\hat{D}} \boldsymbol{\partial} \rk^{- 1} 
\right\rangle_G
\simeq \left\langle \lk - \boldsymbol{\hat{D}} \boldsymbol{\partial} \rk^{- 1} \right\rangle_{\textsc{G}} \lk - \boldsymbol{\partial}^2 \rk 
\left\langle \lk - \boldsymbol{\hat{D}} \boldsymbol{\partial} 
\rk^{- 1}
\right\rangle_{\textsc{G}}
\end{align}
\end{widetext}
to the Coulomb potential, Eqs.~(\ref{384}), (\ref{237}), one finds from the IR analysis of the
 equations of motion of the pure Yang-Mills sector (see e.g.
Ref.~\cite{Heffner:2012sx}) the following relation
\begin{align}
\label{num-1241}
\sigma_{\textsc{C}} = \frac{\pi}{N_{\textsc{C}}} M^2 \, .
\end{align}
For $N_{\textsc{C}} = 3$ we can put $\pi/N_{\textsc{C}} \simeq 1$ and obtain the approximate relation
\begin{align}
\label{num-1246}
\sigma_{\textsc{C}} \simeq M_{\textsc{G}}^2 \, .
\end{align}
With the lattice result $M_{\textsc{G}} \simeq 2 \sqrt{\sigma_{\textsc{W}}}$ this yields $\sigma_{\textsc{C}} \simeq 4 \sigma_{\textsc{W}}$,
which shows that $\sigma_{\textsc{C}}$ is larger than $\sigma_{\textsc{W}}$, in agreement with Ref.~\cite{Zwanziger:2002sh}.

\section{The quark vacuum wave functional}
\label{Gen-functional-sec}

In this section we define our trial state for the quark vacuum $|
\phi_{\textsc{F}} (\vA) \rangle$. For this purpose we decompose
the fermion field $\psi(\bx)$ into positive and negative energy
components
\begin{align}
\label{419-G10}
\psi (\vx) = \psi_+ (\vx) + \psi_- (\vx) \,
\end{align}
given by
\begin{align}
\label{psi+}
 \psi_{\pm}(\bx) &= \int d^3 y \, \Lambda_{\pm}(\bx,\by) \, \psi(\by) \; , \\
\label{psi-}
 \psi^{\dagger}_{\pm}(\bx) &= \int d^3 y \, \psi^{\dagger}(\by) \, \Lambda_{\pm}(\by,\bx)  \;
 ,
\end{align}
where
\begin{align}
\label{Lambda} \Lambda_{\pm}(\bx,\by)= \int \frac{d^3 p}{(2\pi)^3}
\, e^{i\bp \cdot (\bx-\by)} \Lambda_{\pm}(\bp) 
\end{align}
with 
\begin{align}
\Lambda_{\pm}(\bp) =
\frac{1}{2}\left(\mathds{1}\pm\frac{h(\bp)}{E(\bp)} \right) \;
\end{align}
are the projectors onto positive (negative) energy eigenstates.
Here $h(\bp)$ is the free Dirac Hamiltonian
 in momentum
space
\begin{align}
 h(\bp) = \boldsymbol{\alpha}  \bp \,
+ \beta m_{0} \; ,
\end{align}
whose eigenvalues are $\pm E (\bp)$ with $E (\bp) = \sqrt{\vp^2 +
m^2_0}$. The orthogonal projectors fulfill the relations
\begin{align}
\label{436-42}
\Lambda^2_{\pm} = \Lambda_{\pm}, \qquad \Lambda_{\pm}\Lambda_{\mp}=0 , \qquad \Lambda_{\pm} + \Lambda_{\mp} = \mathds{1} .
\end{align}
From the anti-commutation
relation, Eq.~(\ref{172}), the following non-vanishing anti-commutation
relations for the positive (negative) energy spinors follow
\begin{align}
\label{antia} \{\psi_\pm(\bx),\psi_\pm^{\dagger}(\by) \} &=
\Lambda_\pm(\bx,\by) .
\end{align}
The free (bare) fermion vacuum $| 0 \rangle$ is defined by
\begin{align}
\psi_+(\bx) | 0 \rangle = 0 , \qquad \quad \psi_-^{\dagger}(\bx)| 0 \rangle = 0 \; .
\label{perturbative_vacuum}
\end{align}
We choose our trial state $| \phi_{\textsc{F}} \rangle$ of the
quark vacuum as the most general Slater determinant which is not
orthogonal to the bare vacuum $| 0 \rangle$. Such a state has the
form
\begin{align}
\label{coordinate-wave} | \phi_{\textsc{F}} \rangle \!=
\! \mathcal{N}_{\textsc{F}} \exp \left[\!- \!\int \! d^3 x  \int \! d^3 y
\psi^{\dagger}_{+} (\bx) K (\bx, \by) \psi_- (\by) \! \right]
\Big| 0 \Big\rangle \, ,
\end{align}
where $\cN_{\textsc{F}}$ is a normalization constant to be
determined later. The use of a Slater determinant has the
advantage that Wick's theorem applies, which facilitates the
evaluation of expectation values of products of fermion operators.
Since the wave functional
 (\ref{coordinate-wave}) has
to embody the coupling of the quarks to the gluons the kernel $K
(\vx, \vy)$ can in principle be any functional of the gauge field.
We will assume here that $K(\bx,\by)$ can be Taylor expanded in
powers of the gauge field and that this expansion can be truncated
in leading order
\begin{align}
\label{485}
K (\vx, \vy) &= K_0 (\vx, \vy) + \int d^3 z \, \vK (\vx, \vy; \vz) \vA (\vz) \nonumber\\
&\equiv K_0 (\vx, \vy) + K_1 (\vx, \vy)  \, .
\end{align}
From the definition of the wave functional (\ref{coordinate-wave}) and the projection properties (\ref{436-42}) follows that
the variational kernel has the property
\begin{align}
 \int d^3
z \int d^3 z' \Lambda_+(\bx,\bz) K^{a
b}(\bz,\bz') \Lambda_-(\bz',\by) =  K^{a b}(\bx,\by) \; .
\end{align}
Incorporating this property we choose the variational kernels in the form
\begin{widetext}
\begin{align}
\label{K0}
K_0  (\vx, \vy) &= \int d^3 x' \int d^3 y' \Lambda_+ (\vx, \vx') \beta S (\vx' - \vy') \Lambda_- (\vy', \vy)  \\
\label{497-z1}
\vK^a (\vx, \vy; \bz) &= \int d^3 x' \int d^3 y' \Lambda_+ (\vx, \vx') \, \vec{\alpha} \, T^a \, V (\vx' - \vy', \vz - \vy') \Lambda_- (\vy', \vy)  \, ,
\end{align}
\end{widetext}
where the form factors $S (\vx)$ and $V (\vx, \vy)$ are the variational functions to be determined by minimizing the energy density. The choice
of the position arguments in the variational functions is dictated by translational invariance.
Furthermore the Lorentz and color structure of the variational kernel (\ref{497-z1}) is basically dictated by Lorentz and color symmetry since the vacuum wave function
has to be a color and Lorentz scalar. Of course, more complicated (tensor structures and) ansatzes in the exponent are possible
but the present one can be considered as the leading non-trivial order of the expansion of the exponent of the wave functional in powers of the
gauge field.

For $V (\vx, \vy) = 0$ the wave functional $| \phi_{\textsc{F}} \rangle$, Eq.~(\ref{coordinate-wave}),
with the kernel $K$, Eqs.~(\ref{485}), (\ref{K0}), (\ref{497-z1}), reduces to the
BCS-state considered in
Refs.~\cite{Finger:1981gm,LeYaouanc:1983iy,Adler:1984ri}. The new
element is the vector coupling $\vK(\bx,\by) \sim V (\vx, \vy)$,
Eq.~(\ref{497-z1}).

For the explicit calculation it is convenient to express the variational
kernel $K^{a b}(\bx,\by)$ in Eq.~(\ref{coordinate-wave}) in momentum space ($\dbar^3 p = \frac{d^3 p}{(2\pi)^3}$)
\begin{align}
\label{coord-K0}
 K_0 (\bx,\by) &= \int \dbar^3 p \, e^{i \bp \cdot (\bx-\by)} \Lambda_+(\bp) \beta S(\bp) \Lambda_-(\bp) \; , \\
 \label{coord-K1}
 \boldsymbol{K}^a(\bx,\by;\bz) &= \int \dbar^3 p \, \,  \dbar^3 q \, e^{i \bp \cdot (\bx-\by)} e^{i \bq \cdot (\bz-\by)} \times \\
&\qquad  \times \Lambda_+(\bp) \boldsymbol{\alpha} \, T^a \, V( \bp,\bp+\bq) \Lambda_-(\bp+\bq) \; .
\nonumber
\end{align}
The adjoint kernels read
\begin{align}
 K^{\dagger}_0(\bx,\by) &= \int \dbar^3 p \, e^{i \bp \cdot (\bx-\by)} \Lambda_-(\bp) \beta S^{\ast}(\bp) \Lambda_+(\bp) \; , \\
\label{adj-coord-K1} \boldsymbol{K}^{\dagger a}(\bx,\by;\bz) &=
\int \dbar^3 p \, \, \dbar^3 q \, e^{i \bp \cdot (\bx-\by)}
e^{-i
\bq \cdot (\bz-\bx)} \times \\
&\qquad \times \Lambda_-(\bp+\bq) \boldsymbol{\alpha} \, T^a \, V^{\ast}(\bp+\bq, \bp)
\Lambda_+(\bp) \; ,
\nonumber
\end{align}
where $\ast$ means complex conjugation.
The scalar variational function $S(\bp)$ is dimensionless while the vector form factor
 $V(\bp,\bq)$ has dimension of inverse momentum.

\section{The quark generating functional}
\label{Gen-functional-secII}
For the evaluation of the expectation values of quark observables it is convenient to introduce the fermionic generating functional
\begin{widetext}
\begin{align}
\label{531} Z_{\textsc{F}} [\eta] = \langle \phi_{\textsc{F}} |
\exp \left[ \int \lk \eta^*_+ \psi_+ + \eta_- \psi^\dagger_- \rk
\right] \exp \left[ \int \lk \psi^\dagger_+ \eta_+ + \psi_-
\eta^*_- \rk \right] | \phi_{\textsc{F}} \rangle \, ,
\end{align}
\end{widetext}
where $| \phi_{\textsc{F}} \rangle$ is the quark vacuum state
(\ref{coordinate-wave}) and $\eta_+, \eta_-$ are the quark sources,
which are Grassmann valued Dirac spinors. Since $|
\phi_{\textsc{F}} \rangle$ is a
 Slater determinant the
generating functional can evaluated in closed form. One finds after straightforward calculation
\begin{align}
\label{539} Z_{\textsc{F}} [\eta] = |{\cN}_{\textsc{F}} |^2 \, 
\mbox{Det}[ \Omega] \, \exp \left[ \eta^\dagger \Omega^{- 1} \eta
\right] \, ,
\end{align}
where we have introduced the bi-spinor notation
\begin{align}
\label{544}
\eta = \lk \eta_+ \atop - \eta_- \rk
\end{align}
with the matrix $\Omega$ defined by
\begin{align}
\label{549}
\Omega = \lk \begin{array}{cc} \Id & K \\ K^\dagger & - \Id \end{array} \rk \, .
\end{align}
Here $\Id$ denotes the unit kernel in the $+ \atop (-)$ subspace of positive (negative) energy eigenstates.
For vanishing source $\eta$ we find from Eq.~(\ref{539})
\begin{align}
\label{557} Z_{\textsc{F}} [\eta = 0] \equiv \langle
\phi_{\textsc{F}} | \phi_{\textsc{F}} \rangle = |
{\cN}_{\textsc{F}} |^2 \mbox{Det}[\Omega] \, .
\end{align}
Note the norm of $| \phi_{\textsc{F}} \rangle$ is in principle a functional
of the transverse gauge field $\vA$ through the kernel $K$,
Eq.~(\ref{485}). In a fully unquenched calculation only the total
QCD wave functional (\ref{270-y1}) can be normalized. However, in
the quenched calculation the Yang-Mills part and the fermionic
part can be separately normalized. For a quenched calculation we
choose the normalization $\langle \phi_{\textsc{F}} | \phi_{\textsc{F}} \rangle = 1$,
which removes the fermion determinant $\mbox{Det}[\Omega]$ from
the generating functional (\ref{539})
\begin{align}
\label{568} Z_{\textsc{F}} [\eta] = \exp \left[ \eta^\dagger
\Omega^{- 1} \eta \right] \, .
\end{align}
This equation is a compact form of Wick's theorem and allows us to
express all fermionic expectation values in terms of the matrix
$\Omega^{- 1}$. Its gluonic expectation value $\langle \Omega^{-
1} \rangle_{\textsc{G}}$ is closely related to the quark
propagator, see Eq.~(\ref{G17}) below.

The matrix $\Omega$, Eq.~(\ref{549}), can be explicitly inverted
yielding
\begin{eqnarray}
  \label{O-1}
  \Omega^{-1} = \left(
    \begin{array}[c]{cc}
    \left[ \mathds{1}+K K^{\dagger} \right]^{-1}  & \left[\mathds{1}+K K^{\dagger} \right]^{-1} K  \\
\left[\mathds{1}+K^{\dagger} K \right]^{-1} K^{\dagger} & -
\left[\mathds{1}+K^{\dagger} K \right]^{-1}
    \end{array} \right) \, .
\end{eqnarray}
Resolving the bi-spinor structure (\ref{544}) the fermion generating functional (\ref{568}) becomes
\begin{align}
\label{generating-result-split}
Z_{\textsc{F}} \! = \! \exp\Big(\!&\eta^{\ast}_+ \left[ \!\mathds{1}+K K^{\dagger} \right]^{-1} \eta_+ - \eta^{\ast}_- \left[\!\mathds{1}+K^{\dagger} K \right]^{-1} K^{\dagger} \eta_+
\nonumber \\ - &\eta^{\ast}_+ \left[\!\mathds{1}+K K^{\dagger} \right]^{-1} K \eta_- - \eta_-^{\ast}  \left[\!\mathds{1}+K^{\dagger} K \right]^{-1} \eta_-\Big) .
\end{align}
Note that the matrix $\Omega$, Eq.~(\ref{549}), (and hence also
$\Omega^{- 1}$, Eq.~(\ref{O-1})) is overall Hermitian.

With the explicit form of $\Omega^{- 1}$ at hand from Eq.~(\ref{568}) or Eq.~(\ref{generating-result-split}) all fermionic correlation
functions can be evaluated. From the form of the generating
functional (\ref{568}) follows that all fermionic correlation functions can be expressed in terms of the two-point functions, which is a manifestation
of Wick's theorem. For later use we list the non-vanishing two-point functions
\begin{widetext}
\begin{subequations}
\begin{align}
\label{a} \langle \psi_+^a(\bx) \psi_+^{\dagger b}(\by) \,
\rangle_{\textsc{F}} &=
- \frac{\delta^2 Z[\eta]}{\delta \eta_+^{\ast a}(\bx) \, \delta \eta_+^b(\by)} \Big|_{\eta=0} \, =
\left( \Lambda_+ \left[\mathds{1}+K K^{\dagger} \right]^{-1} \Lambda+ \right)^{ab}(\bx,\by) \, , \\
\label{b} \langle \psi_-^{\dagger a}(\bx) \psi_-^{b}(\by) \,
\rangle_{\textsc{F}}  &= - \frac{\delta^2 Z[\eta]}{\delta
\eta_-^{a}(\bx) \, \delta \eta_-^{\ast b}(\by)} \Big|_{\eta=0} =
\left(\Lambda_- \left[\mathds{1} + K^{\dagger} K \right]^{-1} \Lambda_- \right)^{b a}(\by,\bx) \, , \\
\label{atbt} \langle \psi_-^{a}(\bx) \psi_+^{\dagger b}(\by)  \,
\rangle_{\textsc{F}} &=
\frac{\delta^2 Z[\eta]}{\delta \eta_-^{\ast a}(\bx) \delta \eta_+^b(\by)} \Big|_{\eta=0} =
\left(\Lambda_- \left[\mathds{1}+ K^{\dagger} K \right]^{-1} K^{\dagger} \Lambda_+ \right)^{a b}(\bx,\by) \, , \\
\label{ab} \langle \psi_+^a(\bx) \psi_-^{\dagger b}(\by)
\rangle_{\textsc{F}} &= \frac{\delta^2 Z[\eta]}{\delta
\eta_+^{\ast a}(\bx)\, \delta \eta_-^{b}(\by)} \Big|_{\eta=0} =
\left( \Lambda_+ \left[\mathds{1}+ K K^{\dagger} \right]^{-1} K \Lambda_- \right)^{a
b}(\bx,\by) \; ,
\end{align}
\end{subequations}
\end{widetext}
where
the subscript $\textsc{F}$
 denotes the fermion expectation value in the state $| \phi_{\textsc{F}} \rangle$, Eq.~(\ref{coordinate-wave}).
Using the anti-commutation relations (\ref{antia}) we obtain from
Eqs.~(\ref{a}), (\ref{b})
\begin{subequations}
\begin{align}
 \label{at}
\langle  \psi_+^{\dagger a}(\bx) \psi_+^{b}(\by) \rangle_{\textsc{F}}  &=  
\left(\Lambda_+ \left[\mathds{1}+ K K^{\dagger}\right]^{-1} K K^{\dagger}\Lambda_+ \right)^{b a}(\by,\bx) , \\
\label{bt}
\langle  \psi_-^{a}(\bx) \psi_-^{\dagger b}(\by)  \rangle_{\textsc{F}} &=  
\left(\Lambda_- \left[\mathds{1}+K^{\dagger} K \right]^{-1} K^{\dagger} K \Lambda_- \right)^{a b}(\bx,\by) .
\end{align}
\end{subequations}
Through the kernel $K$ the fermionic expectation values $\langle
\ldots \rangle_{\textsc{F}}$ are still functionals of the
transverse gauge field. To find the true correlation functions we
have still to take the gluonic vacuum expectation value of the
fermionic averages $\langle \ldots \rangle_{\textsc{F}}$.
Fortunately, for the Yang-Mills wave functional
(\ref{YM-wave-functional}) Wick's theorem applies. Nevertheless
due to presence of the inverse kernels $(1 + K^\dagger K)^{- 1}$
in the fermionic correlation functions the gluonic expectation
values $\langle \ldots \rangle_{\textsc{G}}$ defined by
Eq.~(\ref{342}) cannot be taken in closed form. To simplify the
calculation we will use the following approximation for inverse
fermionic kernels
\begin{align}
\label{626} \langle \cdots (\Id + K^\dagger K)^{- 1} \ldots
\rangle_{\textsc{G}} \simeq \langle \ldots \lk \Id + \langle
K^\dagger K \rangle_{\textsc{G}} \rk^{- 1} \ldots
\rangle_{\textsc{G}}
\end{align}
i.e.~replacing in the inverse operators the kernels $K^\dagger K$
and $K K^\dagger$ by their expectation values $\langle K^\dagger K
\rangle_{\textsc{G}}$ and $\langle K K^\dagger
\rangle_{\textsc{G}}$, respectively. For the Yang-Mills wave
functional (\ref{YM-wave-functional}) one finds with the explicit
form of $K$, Eq.~(\ref{485}), 
\begin{align}
\begin{split}
\label{638} \langle K^\dagger K \rangle_{\textsc{G}} & =
K^\dagger_0 K_0 + \langle K^\dagger_1 K_1 \rangle_{\textsc{G}} \, , \\
\langle K K^\dagger \rangle_{\textsc{G}} &= K_0
K^\dagger_0 + \langle K_1 K^\dagger_1 \rangle_{\textsc{G}}
\end{split}
\end{align}
with
\begin{widetext}
\begin{align}
 \langle (K^{\dagger}_1 K_1)_{m n}(\bx,\by)\rangle_{\textsc{G}} &= \int d^3 z \int d^3 z' (K_i^{\dagger a})_{m l}
(\bx,\bx';\bz) (K_j^b)_{l n}(\bx',\by;\bz') D_{i j}^{a b}(\bz,\bz') \; , \\
\langle (K_1 K^{\dagger}_1)_{m n}(\bx,\by)\rangle_{\textsc{G}} &= \int d^3 z \int d^3 z' (K_i^{a})_{m l}
(\bx,\bx';\bz) (K_j^{\dagger b})_{l n}(\bx',\by;\bz') D_{i j}^{a b}(\bz,\bz') \; ,
\end{align}
\end{widetext}
where $D^{ab}_{ij} (\vz, \vz')$ is the gluon propagator
(\ref{gluon-prop}). Here we have used that $\langle K_0^{\dagger}
K_1\rangle_{\textsc{G}} = 0 = \langle K_1 K_0^{\dagger}
\rangle_{\textsc{G}}$, since expectation values of an odd number
of gluon fields vanish in the Gaussian vacuum,
Eq.~(\ref{YM-wave-functional}).

For the study of spontaneous breaking of chiral symmetry the small
current quark mass is irrelevant. Therefore from now on we will
put $m_0 = 0$, which will simplify the explicit calculations and,
in particular, the form of the projectors (\ref{Lambda}), which
then satisfy the relation $\beta \Lambda_{-}(\bp)=\Lambda_+(\bp)
\beta$. With the explicit form of the kernels $K_0$,
Eq.~(\ref{coord-K0}), and $\boldsymbol{K}$, Eq.~(\ref{coord-K1}),
one finds after straightforward calculation
\begin{subequations}
\begin{align}
  (K^{\dagger}_0 K_0)(\bx,\by) &= \int \dbar^3 p \, e^{i \bp \cdot (\bx-\by)} S^{\ast}(\bp) S(\bp) \Lambda_{+}(\bp) \; , \\
 (K_0 K^{\dagger}_0)(\bx,\by) &= \int \dbar^3 p \, e^{i \bp \cdot (\bx-\by)} S(\bp) S^{\ast}(\bp)  \Lambda_{-}(\bp) \; , \\
 \langle (K^{\dagger}_1 K_1)_{m n}(\bx,\by)\rangle_{\textsc{G}} &= \delta_{m n} \int \dbar^3 p \, e^{i \bp \cdot (\bx-\by)}  \, R(\bp) \Lambda_+(\bp) \; , \\
 \langle (K_1 K^{\dagger}_1)_{m n}(\bx,\by)\rangle_{\textsc{G}} 
&= \delta_{m n} \int \dbar^3 p \, e^{i \bp \cdot (\bx-\by)}  \, R(\bp) \Lambda_-(\bp) \; .
\end{align}
\end{subequations}
Here we have introduced the loop integral
\begin{align}
\label{R-p}
 R(\bp) \!=\! C_F \int \dbar^3 q  V(\bp,\bq) V^{\ast}(\bq,\bp) D(\boldsymbol{\ell}) 
\left[ 1+ (\hat{\bp} \cdot \hat{\boldsymbol{\ell}})  (\hat{\bq} \cdot 
\hat{\boldsymbol{\ell}}) \right]  ,
\end{align}
where $\boldsymbol{\ell} = \bp - \bq$. Furthermore
\begin{align}
\label{795-16}
D(\boldsymbol{\ell})=
1/(2\omega(\boldsymbol{\ell}))
\end{align}
 is the Fourier-transform of the spatial gluon propagator,
Eq.~(\ref{gluon-prop}), and
$C_F=(N_{\textsc{C}}^2-1)/(2N_{\textsc{C}})$ arises from the
quadratic Casimir.

\section{The quark propagator}
\label{QuarkProp}
To investigate the properties of the quarks in the correlated QCD vacuum the quantity of central interest is the (static) quark propagator
\begin{align}
\label{696}
G_{rs} (\vx, \vy) = \left\langle \phi \left| \frac{1}{2} \left[ \psi_r (\vx), \psi^\dagger_s (\vy) \right] \right| \phi \right\rangle \, .
\end{align}
Working out the fermionic expectation value by means of the generating functional (\ref{531}), (\ref{568}), one finds in the bi-spinor
representation
\begin{align}
\label{G17} G = \langle \Omega^{- 1} \rangle_{\textsc{G}} -
\frac{1}{2} \lk
\begin{array}{cc} \Id & 0 \\ 0 & - \Id \end{array} \rk \, .
\end{align}
To resolve the bi-spinor structure it is more convenient to split
the quark fields in Eq.~(\ref{696}) in its positive and negative
energy components (see Eqs.~(\ref{419-G10}), (\ref{psi+}), (\ref{psi-})) and use
Eqs.~(\ref{a})-({\ref{ab}). This yields
the alternative representation
\begin{widetext}
\begin{align}
G =  \,
&\langle\Lambda_+ (\mathds{1} + K K^{\dagger})^{-1} (\mathds{1} - K K^{\dagger}) \Lambda_+
+\Lambda_- (K^{\dagger} K - \mathds{1}) (\mathds{1} +K^{\dagger} K)^{-1}  \Lambda_- + \nonumber \\ + &\Lambda_- (\mathds{1} + K^{\dagger} K)^{-1} K^{\dagger} \Lambda_+
+ \Lambda_+ (\mathds{1} + K K^{\dagger})^{-1} K \Lambda_-\rangle_{\textsc{G}} \; .
\end{align}
\end{widetext}
Taking now the expectation value in the gluonic Gaussian vacuum,
Eq.~(\ref{YM-wave-functional}), thereby using the approximation
(\ref{626}) and the explicit form of $\langle K^\dagger K
\rangle_{\textsc{G}} , \langle K K^\dagger \rangle_{\textsc{G}}$,
Eq.~(\ref{638}), we eventually obtain for the Fourier transform of
the quark propagator
\begin{widetext}
\begin{align}
\label{Propagator_variational}
G (\vp) =
&\frac{1}{2} \left[\frac{S(\bp)+S^{\ast}(\bp)}{1 + S^{\ast}(\bp)S(\bp) + R(\bp)} \beta +
\frac{1-S^{\ast}(\bp)S(\bp)-R(\bp)}{1 + S^{\ast}(\bp)S(\bp) + R(\bp)} \boldsymbol{\alpha} \hat{\bp} \right] \; .
\end{align}
\end{widetext}
It is interesting to note that the vectorial variational kernel $V(\bp,\bq)$ enters the static quark propagator only via the
loop integral $R(\bp)$, Eq.~(\ref{R-p}).
Setting the vector kernel $V(\vp, \vq)$ to zero, this loop integral vanishes. Due to the approximation
(\ref{626}) the quantity $R (\bp)$, Eq.~(\ref{R-p}), contains the whole effect of the
coupling of the quarks to the transverse spatial gluons.

The quark propagator (\ref{Propagator_variational}) has the expected Dirac structure
\begin{align}
\label{733-z2} G^{- 1} (\vp) = A (\vp) \vec{\alpha} \vp + B
(\vp) \beta = A (\vp) \lk \vec{\alpha} \vp + \beta M (\vp)
\rk \, ,
\end{align}
where
\begin{align}
\label{738}
M (\vp) = \frac{B (\vp)}{A (\vp)}
\end{align}
is the effective quark mass. Inversion of Eq.~(\ref{733-z2})
yields
\begin{align}
\label{743} G (\bp) = \frac{\vec{\alpha} \cdot \vp A (\vp) + \beta B
(\vp)}{\vp^2 A^2 (\bp) + B^2 (\vp)} \, .
\end{align}
Comparing this representation with the explicit form (\ref{Propagator_variational}) we
 obtain the following identifications
\begin{align}
\label{A-dress}
\frac{1}{2} \frac{S(\bp)+S^{\ast}(\bp)}{1+ S^{\ast}(\bp)S(\bp) + R(\bp)} \, = \, \frac{B(\bp)}{B^{2}(\bp) + \vp^2 A^2(\bp)} \; , \\
\label{B-dress}
\frac{1}{2} \frac{1-S^{\ast}(\bp)S(\bp)-R(\bp)}{1 + S^{\ast}(\bp)S(\bp) + R(\bp)} \, = \, \frac{A(\bp) |\vp|}{B^{2}(\bp) + \vp^2 A^2(\bp)} \; .
\end{align}
Dividing Eq.~(\ref{A-dress}) by Eq.~(\ref{B-dress}) we find for the effective quark mass (\ref{738})
\begin{align}
\label{dynam-mass}
M(\bp) \, = \, |\bp| \, \frac{S(\bp)+S^{\ast}(\bp)}{1- S^{\ast}(\bp)S(\bp) -R(\bp)} \; .
\end{align}
For a non-vanishing scalar form factor $S (\vp)$ (i.e. for
non-vanishing quark-antiquark correlations, see
Eqs.~(\ref{coordinate-wave}), (\ref{485}), (\ref{497-z1})) a quark
mass is dynamically generated. This dynamical mass generation is a
consequence of the spontaneous breaking of chiral symmetry, which
is signaled by a non-vanishing quark condensate
\begin{align}
\left\langle  \overline{\psi}^a(\bx) \psi^a(\bx) \right\rangle = -  \, \int \dbar^3 p \, \mbox{tr} \left[
\beta \, G(\bp) \right] \; .
\end{align}
Inserting here the explicit form  of the quark propagator  (\ref{Propagator_variational}) we find
\begin{align}
\label{chiral-cond}
\left\langle  \overline{\psi}^a(\bx) \psi^a(\bx)  \right\rangle = - N_{\textsc{C}} \, 2 \, \int \dbar^3 p \frac{S(\bp) + S^{\ast}(\bp)}{1+S^{\ast}(\bp) S(\bp)+R(\bp)} \; .
\end{align}
Obviously, a non-vanishing quark condensate requires $S (\vp) \neq 0$. Thus a wave functional (\ref{coordinate-wave}) with vector coupling
only $(S (\vp) = 0)$ cannot yield spontaneous breaking of chiral symmetry.
Whether chiral symmetry is spontaneously broken is a dynamical question and requires the determination of the kernels $S (\vp)$ and
$V (\vp, \vq)$ in the quark wave functional (\ref{coordinate-wave}). This will be done in the following
 sections by means of the variational approach.

\section{Energy densities and gap equations}
\label{Dirac-E} We are now in a position to explicitly calculate
the expectation value of the QCD-Hamiltonian. For a quenched
calculation the pure gluonic part $\overline{H}^C_{\textsc{YM}}$,
Eq.~(\ref{300}), can be ignored. We begin with the
Dirac-Hamiltonian $H_{\textsc{F}}$, Eq.~(\ref{167-x4}), whose
expectation value reads
\begin{align}
 \langle H_{\textsc{F}} \rangle = \int d^3 x \int d^3 y \, \left(-i \boldsymbol{\alpha} \boldsymbol{D}\right)_{r s}
\left\langle \psi^{\dagger}_r(\bx) \psi_s(\by) \right\rangle \; ,
\end{align}
where  the covariant derivative is given in Eq.~(\ref{177-x5}).
Splitting the fermion field $\psi(\bx)$ in its positive and
negative energy components, Eq.~(\ref{419-G10}), one observes that
only the expectation values $ \langle \psi_{+}^{\dagger
a}(\bx)\psi^b_{+}(\by) \rangle$ and $\langle \psi^{\dagger
a}_{-}(\bx) \psi_{-}^{b}(\by) \rangle$ contribute to the kinetic
energy of the quarks $\sim \vec{\alpha} \vp$, while the
coupling to the transverse gluons receives contributions from $
\langle \psi_{+}^{\dagger a}(\bx)\psi^b_{-}(\by) \rangle$ and
$\langle \psi^{\dagger a}_{-}(\bx) \psi_{+}^{b}(\by) \rangle$. One
finds
\begin{widetext}
\begin{align}
\begin{split}
 \label{qgc-expect}
&\frac{\langle H_{\textsc{F}} \rangle}{\delta^{3}(0)} = 2 \, N_{\textsc{C}} \, \int d^3 p \,|\bp| \, \frac{S^{\ast}(\bp) S(\bp) + R(\bp)-1}{1+S^{\ast}(\bp) S(\bp)+R(\bp)}  
+  \\
&+2 \, g \, N_{\textsc{C}} \, C_F \, (2\pi)^3 \, \int \dbar^3 p
\int \dbar^3 q \frac{V^{\ast}(\bp,\bq) +
V(\bp,\bq)}{1+S^{\ast}(\bp) S(\bp)+ R(\bp)} D(\boldsymbol{\ell})
\left[1+ (\hat{\bp} \cdot \hat{\boldsymbol{\ell}})(\hat{\bq} \cdot
\hat{\boldsymbol{\ell}})\right]  \; , 
\end{split}
\end{align}
\end{widetext}
where we have set $\boldsymbol{\ell} = \vp - \vq$ and $D (\vp) = 1 /(2 \omega
(\vp))$ is the Fourier transform of the gluon propagator
(\ref{gluon-prop}).

To evaluate the energy density of $H_{\textsc{C}}^{\textsc{F}}$, Eq.~(\ref{H-quark-all}), we replace the Coulomb kernel $ \hat{F}^{a b}(\bx,\by)$,
Eq.~(\ref{237}), by its gluonic expectation value, Eq.~(\ref{384}).
This approximation is consistent with the quenched approximation
 and with the
approximation (\ref{626}) used for the kernels in the denominator
of the quark propagator. One can easily convince oneself that
within our approximation the coupling term
$H^{\textsc{coupl}}_{\textsc{C}} \sim \rho_{\textsc{YM}}
\rho_{\textsc{F}}$ does not contribute, $\langle
H^{\textsc{coupl}}_{\textsc{C}} \rangle = 0$. For this we notice
that for a color diagonal gluon propagator $\langle
\rho_{\textsc{YM}} \rangle_{\textsc{G}} = 0$. Furthermore $\langle
\rho_{\textsc{F}} \rangle_{\textsc{F}}$ contains within our
approximation (\ref{626}}) at most terms linear in the gauge field
$A$. Since $\rho_{\textsc{YM}}$ is quadratic in the gauge field
for the Gaussian Yang-Mills wave functional follows $ \langle
\rho_{\textsc{YM}} \langle \rho_{\textsc{F}} \rangle_{\textsc{F}}
\rangle_{\textsc{G}} = 0 \, . $

The expectation value of the Coulomb Hamiltonian
$H_{\textsc{C}}^{\textsc{F}}$, given by $H_{\textsc{C}}$,
Eq.~(\ref{232}), with total color charge $\rho^a$ replaced by the
quark part $\rho^a_{\textsc{F}}$, Eq.~(\ref{196-x8}), is
straightforwardly evaluated by splitting the fermion fields
$\psi,\psi^{\dagger}$ into their positive and negative energy
components, Eqs.~(\ref{psi+}), (\ref{psi-}), and applying
Wick's theorem. The final form of the Coulomb energy density is
\begin{widetext}
\begin{align}
\label{Coulomb-energy} \frac{\langle H_{\textsc{C}}^{\textsc{F}} \rangle }{\delta^3(0)} \,
= \, \frac{1}{2} N_{\textsc{C}} C_F (2\pi)^3 \int \dbar^3 p \, \,
\dbar^3 q \, V_{\textsc{C}}(\bp-\bq) \, \left[ Y(\bp,\bq) +
Z(\bp,\bq) \hat{\bp} \cdot \hat{\bq} \right] \; ,
\end{align}
\end{widetext}
where we have introduced the abbreviations
\begin{widetext}
\begin{subequations}
\begin{align}
\label{M} Y(\bp,\bq) \, &= \, 1 - \frac{S^{\ast}(\bp) S(\bq) +
S(\bp) S^{\ast}(\bq) + S(\bp) S(\bq)+S^{\ast}(\bp)
S^{\ast}(\bq)}{(1+S^{\ast}(\bp) S(\bp)+ R(\bp))
(1+S^{\ast}(\bq) S(\bq)+ R(\bq))} \; , \\
\label{N} Z(\bp,\bq) \, &= \, - \frac{(1-S^{\ast}(\bp)
S(\bp)-R(\bp)) (1-S^{\ast}(\bq) S(\bq)- R(\bq))}{(1+S^{\ast}(\bp)
S(\bp)+R(\bp)) (1+S^{\ast}(\bq) S(\bq)+R(\bq))} + \nonumber \\ &
\quad \, + \, \frac{-S^{\ast}(\bp) S(\bq) - S(\bp) S^{\ast}(\bq) +
S(\bp) S(\bq)+ S^{\ast}(\bp) S^{\ast}(\bq)}{(1+S^{\ast}(\bp)
S(\bp)+R(\bp)) (1+S^{\ast}(\bq) S(\bq)+R(\bq))} \; .
\end{align}
\end{subequations}
\end{widetext}
Note that $Y (\vp, \vq)$ and $Z (\vp, \vq)$ are both real.
The vector kernel $V$ enters the Coulomb energy density only
through the loop integral $R(\bp)$, Eq.~(\ref{R-p}).
Putting $V (\bp, \bq) = 0$, the loop integral $R(\bp)$
vanishes and the energy density
(\ref{Coulomb-energy}) reduces to the expression obtained in
Ref.~\cite{Adler:1984ri} for the BCS-wave functional.
Since the energy density given by Eqs.~(\ref{qgc-expect}) and (\ref{Coulomb-energy}) is real the
variation with respect to $S$ and $S^{\ast}$ (or $V$ and $V^{\ast}$)
lead to complex conjugate equations, which can be shown to allow for real solutions.
In the following we therefore set $S(\bp)=S^{\ast}(\bp)$,
$V(\bp,\bq)=V^{\ast}(\bp,\bq)$. The second term in $Z(\bp,\bq)$,
Eq.~(\ref{N}), then vanishes.

Minimizing the energy densities $\langle H_{\textsc{F}} \rangle$,
Eq.~(\ref{qgc-expect}), and $\langle H^{\textsc{F}}_{\textsc{C}}
\rangle$, Eq.~(\ref{Coulomb-energy}), with respect to the
variational kernels $S(\bk)$ and $V(\bk,\bk')$ we obtain the
following system of coupled integral equations
\begin{align}
\label{S-ii}
S(\bk) &=  \frac{\frac{1}{2} C_F I^{(1)}_\textsc{C}(\bk)}{|\bk| - \frac{g}{2} C_F I_{\omega}(\bk)} \; , \\
\label{V-i} V(\bk,\bk') &= - g \frac{1+S^2(\bk)+R(\bk)}{2 |\bk| -
g C_F I_{\omega}(\bk) +C_F  \,
I^{(2)}_\textsc{C}(\bk) } \; ,
\end{align}
where we have introduced the loop integrals ($\boldsymbol{\ell}=\bk-\bq$)
\begin{align}
\label{I-omega} I_{\omega}(\bk)  = 2 \int \dbar^3 q  V(\bk,\bq)
D(\boldsymbol{\ell}) \left[ 1+ (\hat{\bk} \cdot
\hat{\boldsymbol{\ell}}) \, (\hat{\bq} \cdot
\hat{\boldsymbol{\ell}}) \right] 
 \; ,
\end{align}
\begin{widetext}
\begin{subequations}
\begin{align}
\label{IC1}
I^{(1)}_\textsc{C}(\bk) \, &= \, \int \dbar^3 q \frac{V_{\textsc{C}}(\bk-\bq)}{1+S^2(\bq)+R(\bq)} 
\left[ S(\bq) (1-S^2(\bk) + R(\bk)) - (\hat{\bk} \cdot
\hat{\bq}) S(\bk) \left(1-S^2(\bq) - R(\bq) \right) \right] \; , \\
\label{IC2} I^{(2)}_\textsc{C}(\bk) \, &= \, \int \dbar^3 q
\frac{V_{\textsc{C}}(\bk-\bq)}{1+S^2(\bq)+R(\bq)} \left[2 S(\bk)
S(\bq) + (\hat{\bk} \cdot \hat{\bq}) \left( 1-S^2(\bq) - R(\bq)
\right) \right] \, .
\end{align}
\end{subequations}
\end{widetext}
The gap equation (\ref{V-i}) determines the variational function
$V (\vk, \vk')$ to depend on its first momentum argument only,
i.e.~\footnote{In an unquenched calculation, taking the variation
of the pure Yang-Mills part $\overline{H}^C_{\textsc{YM}}$,
Eq.~(\ref{300}), into account, this statement no longer holds
true.}
\begin{align}
V (\vk, \vk') = V (\vk) \; .
\end{align}
With this form of the vector kernel
the loop integrals $I_{\omega}(\bk)$,
Eq.~(\ref{I-omega}), and $R(\bk)$, Eq.~(\ref{R-p}), simplify to
\begin{align}
\label{I-bk1} I_{\omega}(\bk) =  2 \,  V(\bk) \, I(\bk) , \qquad
R(\bk) = C_F V^2(\bk) I(\bk) \; ,
\end{align}
with
\begin{align}
\label{I-bk} I(\bk) =  \int \dbar^3 q \, D(\boldsymbol{\ell})
\left[ 1+ (\hat{\bk} \cdot \hat{\boldsymbol{\ell}}) \, (\hat{\bq}
\cdot \hat{\boldsymbol{\ell}}) \right], \qquad
\boldsymbol{\ell}=\bk-\bq \; .
\end{align}
Then the system of coupled equations (\ref{S-ii}), (\ref{V-i})
becomes after the replacement $V (\vk) \to (- V (\vk))$,
~\footnote{By the definition of $V$ (see Eqs.~(\ref{coordinate-wave}), 
(\ref{485}), (\ref{497-z1})) this
replacement is equivalent to changing the sign of the gauge field
$\vA (\vx)$ or of the coupling constant $g$, which leaves the
theory invariant.}
\begin{align}
\label{Gap1}
S(\bk) &=  \frac{\frac{1}{2} C_F I^{(1)}_\textsc{C}(\bk)}{|\bk| + g C_F V(\bk) I(\bk)} \; , \\
\label{Gap2} V(\bk) &=  \frac{g}{2} \frac{1+S^2(\bk)+R(\bk)}{|\bk|
+ g C_F V(\bk) I(\bk) +\frac{1}{2} C_F \,
I^{(2)}_\textsc{C}(\bk) } \; .
\end{align}
Once these equations are solved the quark part of the vacuum wave
functional of QCD is known and all quark observables can, in
principle, be evaluated. These equations need the gluon propagator
(\ref{gluon-prop}) and the non-Abelian Coulomb potential
(\ref{384}) as input. For the gluon propagator we will use the
Gribov formula (\ref{Gribov-fit}). Following
Ref.~\cite{Adler:1984ri} for the Coulomb potential we use the
confining form (\ref{389}), which in momentum space reads
\begin{align}
\label{Ch-960-22}
V_{\textsc{C}}(\vk) = \frac{8 \pi \sigma_{\textsc{C}}}{\vk^4} \, .
\end{align}
With this potential the Coulomb loop integrals
$I^{(1)}_{\textsc{C}}(\bk)$, Eq.~(\ref{IC1}), and
$I^{(2)}_{\textsc{C}}(\bk)$, Eq.~(\ref{IC2}), are UV-finite. Then the
only UV-divergent quantity occurring in the gap equations is the
gluon loop integral $I(\bk)$, Eq.~(\ref{I-bk}).

The gluon energy (\ref{Gribov-fit}) contains a mass scale (Gribov mass) $M_{\textsc{G}}$ which separates 
the UV- and IR-regions of the gluon propagator.
 To isolate the divergencies
of $I(\bk)$ we replace the gluon propagator in the momentum regime $q > M_{\textsc{G}}$ by its
UV-part
\begin{align}
\label{Ch-9622a} D_{\textsc{UV}} (\vk) = 1/(2 |\vk|) \; , 
\end{align}
and define the UV-part of the gluon loop integral as
\begin{widetext}
\begin{align}
\label{GluonUV}
 I_{\textsc{UV}}(\bk) = \int \frac{d^3 q}{(2\pi)^3} \, 
D_{\textsc{UV}}(\bk-\bq) \, \Theta(|\bq| - M_{\textsc{G}})
\left[ 1 + (\hat{\bk} \cdot \hat{\boldsymbol{\ell}}) (\hat{\bq} \cdot \hat{\boldsymbol{\ell}}) \right] \: ,
\qquad \boldsymbol{\ell} = \bk -\bq \: .
\end{align}
\end{widetext}
The $\Theta$-function ensures that only loop momenta $q$ larger than the Gribov mass scale $M_{\textsc{G}}$
contribute. Introducing a momentum cutoff $\Lambda$ this integral is readily evaluated. 
Separating divergent and finite pieces
\begin{align}
I_{\textsc{UV}} (\vk, \Lambda) = I^{\text{fin}}_{\textsc{UV}}(\vk) + I^{\text{div}}_{\textsc{UV}}(\vk, \Lambda)
\end{align}
we find
\begin{align}
\label{1104-21}
 I^{\text{div}}_{\textsc{UV}}(k, \Lambda) = \frac{1}{8 \pi^2} \lk \Lambda^2 - \frac{2}{3}
 \Lambda k \rk \: , 
\end{align}
and
\begin{widetext}
\begin{align}
\label{I-REG-UV}
 I^{\text{fin}}_{\textsc{UV}}(k) = 
\frac{1}{8 \pi^2} \left[ \left(-M^2_{\textsc{G}} + \frac{2}{3} k M_{\textsc{G}} \right) \Theta(M_{\textsc{G}} -k) 
+ \left(-\frac{2}{3} \frac{M_{\textsc{G}}^3}{k} + \frac{1}{6} \frac{M_{\textsc{G}}^4}{k^2} + \frac{1}{6} k^2 \right) \Theta(k - M_{\textsc{G}}) \right] \; . 
\end{align}
\end{widetext}
Note that the UV-divergent part $I^{\text{div}}_{\textsc{UV}}(k)$, Eq.~(\ref{1104-21}), is independent of 
the Gribov mass scale $M_{\textsc{G}}$, which was used to define the UV-regime. A comment is here in order:
A shift of the integration variable $q \to l$ would make the integral a (diverging) constant 
independent of $k$. However, as is well known, shifting the integration variable  before
regularization changes the value of the regularized integral. It is important to keep the 
momentum rooting as in Eq.~(\ref{GluonUV}] for which the UV finite part of the gluon loop integral,
$I^{\text{fin}}_{\textsc{UV}}(k)$, Eq.~(\ref{I-REG-UV}), has the asymptotic behavior
\begin{align}
\label{UV-lead}
 I^{\text{fin}}_{\textsc{UV}}(k \rightarrow \infty) = 
\frac{k^2}{48 \pi^2}  \; ,
\end{align}
which ensures that the dressing function $V (\vk)$ has the correct UV-perturbative behavior, 
as we will see in the next section.

\begin{figure}
\originalTeX
\centering{
\includegraphics[width=.9\linewidth]{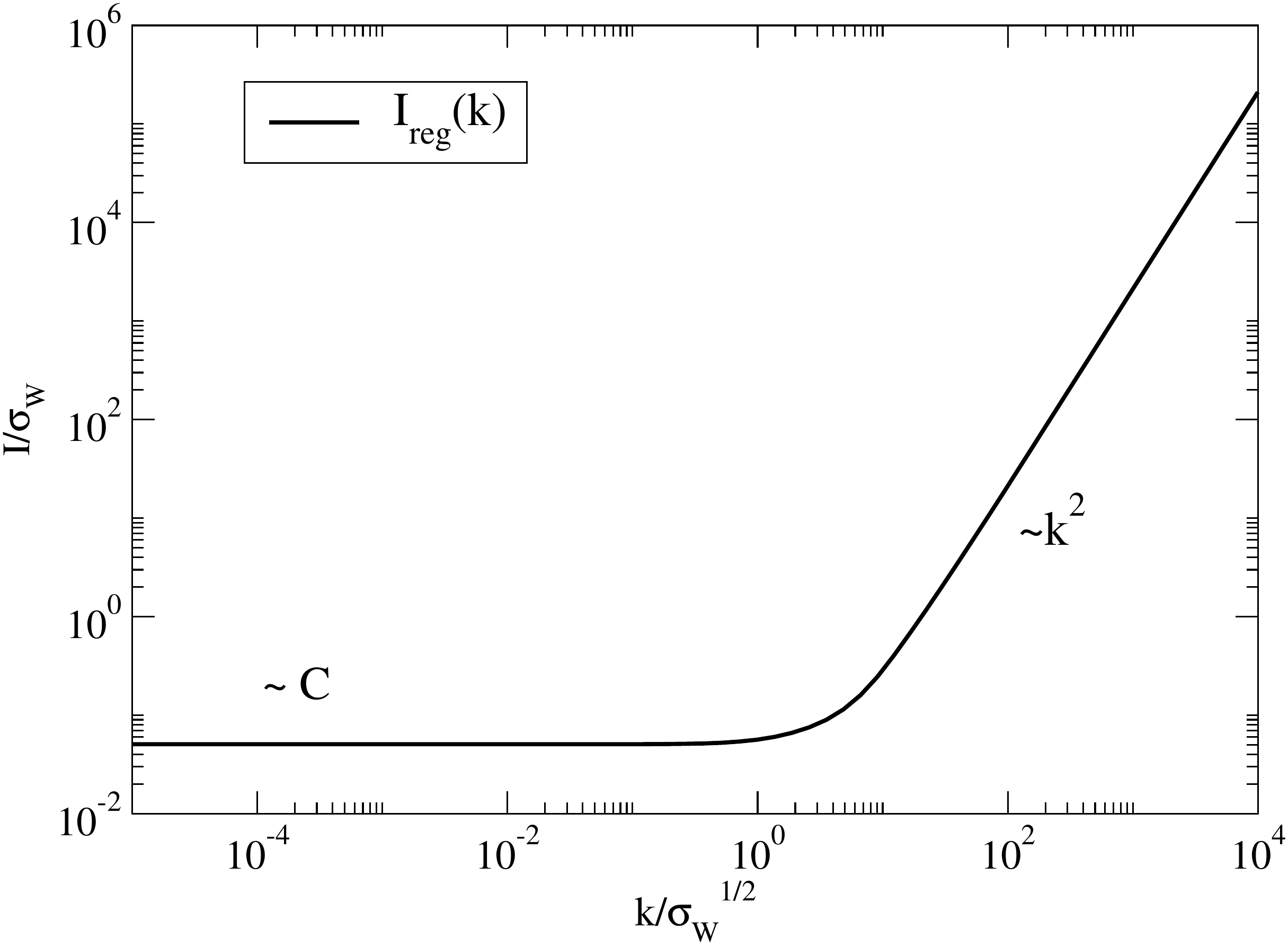}
}
\caption[Gluon loop integral]{\sl Regularized loop integral $I_{\text{reg}} 
(k)$ in units of the string
tension $\sqrt{\sigma_{\textsc{W}}}$, with $M_{\textsc{G}}$ given in Eq.~(\ref{Gribov-Mass}).
}
\label{fig-I-omega}
\end{figure}

We now define the regularized part of the gluon loop integral (\ref{I-bk}) by subtracting 
its UV-divergent piece $I^{\text{div}}_{\textsc{UV}}(\vk, \Lambda)$
\begin{align}
\label{1132-21a}
I_{\text{reg}} (k) = \lim\limits_{\Lambda \to \infty} \left[ I (k, \Lambda) - I^{\text{div}}_{\textsc{UV}} (k, \Lambda)
\right] + C \, ,
\end{align}
where $C$ is an arbitrary finite renormalization constant. In principle, this constant could be 
determined by minimizing the energy density. This would, however, require to renormalize not only the gap equations 
(\ref{Gap1}), (\ref{Gap2}) but also the energy density itself, which is quite involved and 
which we have not done yet. However, we can circumvent this problem by noticing that the quark condensation occurs in order 
to lower the energy of the system. (The superconducting ground state has a lower 
energy than the normal state.) We can therefore assume to minimize the energy density by maximizing  
the quark condensate. Hence, we will choose $C$ so as to maximize the magnitude of the 
quark condensate. Numerically we have found that the optimal value is $C=M^2_{\textsc{G}}/(8\pi^2)$. The regularized gluon loop integral 
(\ref{1132-21a}) is plotted in Fig.~\ref{fig-I-omega}.

\section{Asymptotic analysis}
\label{Asympt-sect}
Below we analyze the coupled equations (\ref{Gap1}), (\ref{Gap2}) in the IR and UV.
For this purpose we analyze first the loop integrals $I^{(1)}_\textsc{C}(\bk)$,
Eq.~(\ref{IC1}), $I^{(2)}_\textsc{C}(\bk)$, Eq.~(\ref{IC2}),
$I(\bk)$, Eq.~(\ref{I-bk}).

The angular parts of the Coulomb integrals $I_{\textsc{C}}^{(1)}(\bk)$ and
$I_{\textsc{C}}^{(2)}(\bk)$ can be reduced to the following two types of
angular integrals
\begin{align}
 \int d \Omega_q \, V_{\textsc{C}}(\bk-\bq) , \qquad \int d \Omega_q \, V_{\textsc{C}}(\bk-\bq) \, \hat{\bk} \cdot \hat{\bq} \; ,
\end{align}
where $\int d \Omega_q = \il^\pi_0 \sin \theta d \theta \il^{2 \pi}_0 d \varphi$ and $\hat{\vk} \cdot \hat{\vq} = \cos \theta$. The
$\varphi$-integral is trivial. For the Coulomb potential (\ref{Ch-960-22}) the integrals over $z = \cos \theta$ can be also
taken analytically using 
\begin{widetext}
\begin{subequations}
\begin{align}
\label{Lp}L (k,q)  &\equiv  \int_{-1}^{1} dz \, \frac{1}{(k^2-2kq z + q^2)^2} \, = \, \frac{2}{(k^2-q^2)^2} \; , \\
\label{Kp}K(k,q) &\equiv  \int_{-1}^{1} dz \, \frac{z}{(k^2-2kq z +
q^2)^2} \, = \, \frac{k^2+q^2}{k q (k^2-q^2)^2} + \frac{1}{2}
\frac{1}{k^2 q^2} \ln\left|\frac{k-q}{k+q} \right| \; .
\end{align}
\end{subequations}
\end{widetext}

\subsection{IR analysis}
We first study the IR-behaviour by expanding
 the kernels $L(k,q)$, Eq.~(\ref{Lp}), and $K(k,q)$,
Eq.~(\ref{Kp}), in power of $k$, yielding
\begin{align}
\label{Lp-small}
L(k,q) &= \frac{2}{q^4} + \frac{4 k^2}{q^6} + \mathcal{O}(k^4) \; , \\
\label{Lp-small-1041}
K(k,q) &= \frac{8}{3}\frac{k}{q^5} + \mathcal{O}(k^3) \; .
\end{align}
For $k\rightarrow 0$ the leading contribution comes from the
kernel $L(k,q)$, while the kernel $K(k,q)$ vanishes at this order. For
vanishing momenta the (regularized) integral $I(k)$,
Eq.~(\ref{1132-21a}), is IR-finite. Assuming the variational functions
$S(k), V(k)$ to be analytic in the IR-region the gap equation (\ref{Gap1}) for $S (k)$ reduces for $k \to 0$ to
\begin{align}
\label{Gap1-infrared}
g C_F V(0) I(0)  \, S(0)
= \widetilde{I}_C^{(1)}(0) \, ,
\end{align}
where
\begin{align}
 \widetilde{I}_C^{(1)}(0) = \left[ 1- S^2(0) + C_F
V^2(0) I(0) \right]  \,  I^{\textsc{IR}}_{\textsc{C}}(0)
\end{align}
 and we have introduced the abbreviation \footnote{We note that the integral $I^{\textsc{IR}}_{\textsc{C}}(0)$ is finite when 
 the infrared regulator $\varepsilon$, Eq.~(\ref{num,-1261}) is introduced, which is done in the numerical evaluation.}
\begin{align}
\label{I-IR1}
  I^{\textsc{IR}}_{\textsc{C}}(0) = G \int dq q^2 \, L(k=0,q) \frac{S(q)}{1+S^2(q)+R(q)}  \; ,
\end{align}
with $L(k,q)$ given in Eq.~(\ref{Lp-small}) and $G = \sigma_{\textsc{C}}/\pi$.
Eq.~(\ref{Gap1-infrared}) is a quadratic equation and can be solved as
\begin{widetext}
\begin{align}
\label{S-infrared}
 S(0)  = \frac{-g C_F V(0) I(0)  \pm \sqrt{\left(g C_F V(0) I(0)\right)^2
 + 4 \left(I^{\textsc{IR}}_{\textsc{C}}(0)\right)^2 (1 + C_F V^2(0) I(0) )}}{2 I^{\textsc{IR}}_{\textsc{C}}(0)} .
\end{align}
\end{widetext}
For the BCS-type of wave functional, defined by
Eq.~(\ref{coordinate-wave}) with vanishing vector kernel $V(k)$,
Eq.~(\ref{Gap1-infrared}) simplifies to
\begin{align}
 0 = \left[ 1- S^2(0) \right] I^{\textsc{IR}}_{\textsc{C}}(0)
\end{align}
and is solved for $S(k\rightarrow 0) = \pm 1$. This solution is of course also obtained from Eq.~(\ref{S-infrared}) with $V (0) = 0$.
With the
coupling of the quarks to the
transverse gluons included $(V (k) \neq 0)$ \textit{all} parts of the QCD energy contribute to
 the infrared value $S (0)$ of the scalar gap function
$S(k)$, which is then no longer constrained to $\pm 1$
\footnote{However, using a perturbative gluon propagator, i.e.
$\omega_{\textsc{UV}}(\bk)=|\bk|$, the infrared value of $S$
is, as for the BCS-case, constrained to be unity (since the loop
integral $I(k)$, Eq.~(\ref{I-bk}), then vanishes like  $k^2$ for
small momenta).}.

To find the infrared value of $V(k)$, we take the $k \to 0$ limit of
Eq.~(\ref{Gap2}), which yields
\begin{widetext}
\begin{align}
\label{CH-1088-ae}
 \left[ g C_F V( 0) I( 0) +  \widetilde{I}_C^{(2)}(0) \right] V(0)
= \frac{g}{2} \left[ 1+ S^2(0) + C_F V^2(0) I(0) \right] \; ,
\end{align}
\end{widetext}
with
\begin{align}
 \widetilde{I}_C^{(2)}(0) = 2 S(0) \, I^{\textsc{IR}}_{\textsc{C}}(0)
\end{align}
and $I^{\textsc{IR}}_{\textsc{C}} (0)$ defined by Eq.~(\ref{I-IR1}). Eq.~(\ref{CH-1088-ae}) can be solved for $V (0)$ yielding
\begin{widetext}
\begin{align}
\label{V-infrared}
V(0) = \frac{-2S(0) I^{\textsc{IR}}_{\textsc{C}} \pm \sqrt{(2S(0) I^{\textsc{IR}}_{\textsc{C}})^2 +
(g C_F I(0)) (g (1+S^2(0)))}}{g C_F
I(0)} \; .
\end{align}
\end{widetext}
Like $S (k)$, the vector kernel $V (k)$ is IR-finite. Since $V (k =  0) \neq 0$
 we can expect, that the coupling of the quarks
to the transverse gluons is indeed relevant for the infrared physics.

\subsection{UV-analysis}

Due to asymptotic freedom we expect the coupling kernels $S (\vk)$ and $V (\vk)$ to vanish for $k \to \infty$. Therefore we
make the following power-law ansatzes
\begin{align}
\label{power}
 S(k\rightarrow \infty) = \frac{A}{k^{\alpha}} , \qquad V(k\rightarrow \infty) = \frac{B}{k^{\beta}} \; .
\end{align}
Since the integrals $L (k, q), K (k, q)$, Eqs.~(\ref{Lp}),
(\ref{Kp}), are symmetric in the two entries $k$ and $q$ the
UV-behaviour of these integrals for large $k \to \infty$ can be
obtained from the IR-expressions for $k \to 0$,
Eqs.~(\ref{Lp-small}), (\ref{Lp-small-1041}), by interchanging the
momenta $k \leftrightarrow q$, yielding
\begin{align}
\label{Lp-UV}
L(k,q) &= \frac{2}{k^4} + \mathcal{O}\left(\frac{1}{k^6}\right) \; , \\
K(k,q) &= \frac{8 q}{3 k^5} +
\mathcal{O}\left(\frac{1}{k^7}\right) \; .
\end{align}
With these expressions one finds for the UV-behavior of the Coulomb integrals
(\ref{I-omega}), (\ref{IC1})
\begin{widetext}
\begin{subequations}
\begin{align}
\label{CH1127-ar} I^{(1)}_{\textsc{C}} (k \to \infty) &= \lk
\frac{2}{k^4} + O \lk \frac{1}{k^6} \rk \rk \left[ 1 - S (k) + C_F
V^2 (k) I (k) \right] I^{\textsc{UV}}_{\textsc{C}}
\nonumber\\
I^{(2)}_{\textsc{C}} (k \to \infty) &=  \lk \frac{2}{k^4} + O \lk
\frac{1}{k^6} \rk \rk 2 S(k) I^{\textsc{UV}}_{\textsc{C}} \, ,
\end{align}
\end{subequations}
\end{widetext}
with \begin{align} I^{\textsc{UV}}_{\textsc{C}} = G \int dq \, q^2
\, \frac{S(q)}{1+S^2(q) + R(q)} \; .
\end{align} Furthermore the finite
 UV leading term of the gluon loop integral $I(k)$, Eq.~(\ref{I-omega}), is given 
by $I^{\text{fin}}_{\textsc{UV}} (k) \sim k^2$,
 see Eq.~(\ref{UV-lead}).

With the UV-behavior of the loop integrals at hand it is now straightforward to carry out the UV-analysis of the gap equations
(\ref{I-bk}), (\ref{Gap1}). One finds the following behavior
\begin{align}
\label{CH-1137-aw}
S (k) \sim 1/k^5 \, , \quad \quad V (k) \sim 1/k \, , \quad \quad k \to \infty \, .
\end{align}
The same UV-behaviour of $S (k)$ was found in
Ref.~\cite{Adler:1984ri}.

The results obtained in this chapter in the IR- and UV-analysis are all confirmed in the numerical solution of the gap
equations (\ref{Gap1}), (\ref{Gap2}). The variational functions
show a perfect power law behavior for large momenta, see
Fig.~\ref{fig-s-v-asymptotic}.

\section{Chiral Symmetry Breaking with spatial gluons only?}\label{sectionIX}

We first explore whether, neglecting the Coulomb potential
$V_{\textsc{C}}(\bk)$, Eq.~(\ref{Ch-960-22}), the coupling of the quarks to the transverse
gluons alone can generate spontaneous breaking of chiral symmetry.

Neglecting the Coulomb potential implies $I^{(1, 2)}_{\textsc{C}}
(k) = 0$, Eqs.~(\ref{IC1}), (\ref{IC2}) and simplifies the equations of motions (\ref{Gap1}),
(\ref{Gap2}) to
\begin{align}
\label{CH-1158-g1}
S (k) \lk k + g C_F V(k) I (k) \rk & =  0 \\
\label{CH-1160-g2} V (k) &= \frac{g}{2} \frac{1 + S^2 (k) + C_F
V^2 (k) I (k)}{k + g C_F V (k) I (k)} \, .
\end{align}
A non-vanishing quark condensate (\ref{chiral-cond}) requires  $S (k) \neq 0$, for which equation (\ref{CH-1158-g1}) reduces to
\begin{align}
\label{CVH-1165} k + g C_F V(k) I (k) = 0 \, .
\end{align}
This equation has no solution, in particular for $k = 0$, since $I (k = 0) \neq 0$. Hence, with the neglect of the color Coulomb
potential $V_{\textsc{C}}(\bk)$, Eq.~(\ref{Ch-960-22}), only the trivial solution $S (k) = 0$ exists, i.e. spontaneous breaking
of chiral symmetry does not occur.

\begin{figure}
\originalTeX \centering{
\includegraphics[angle=-90,width=.9\linewidth]{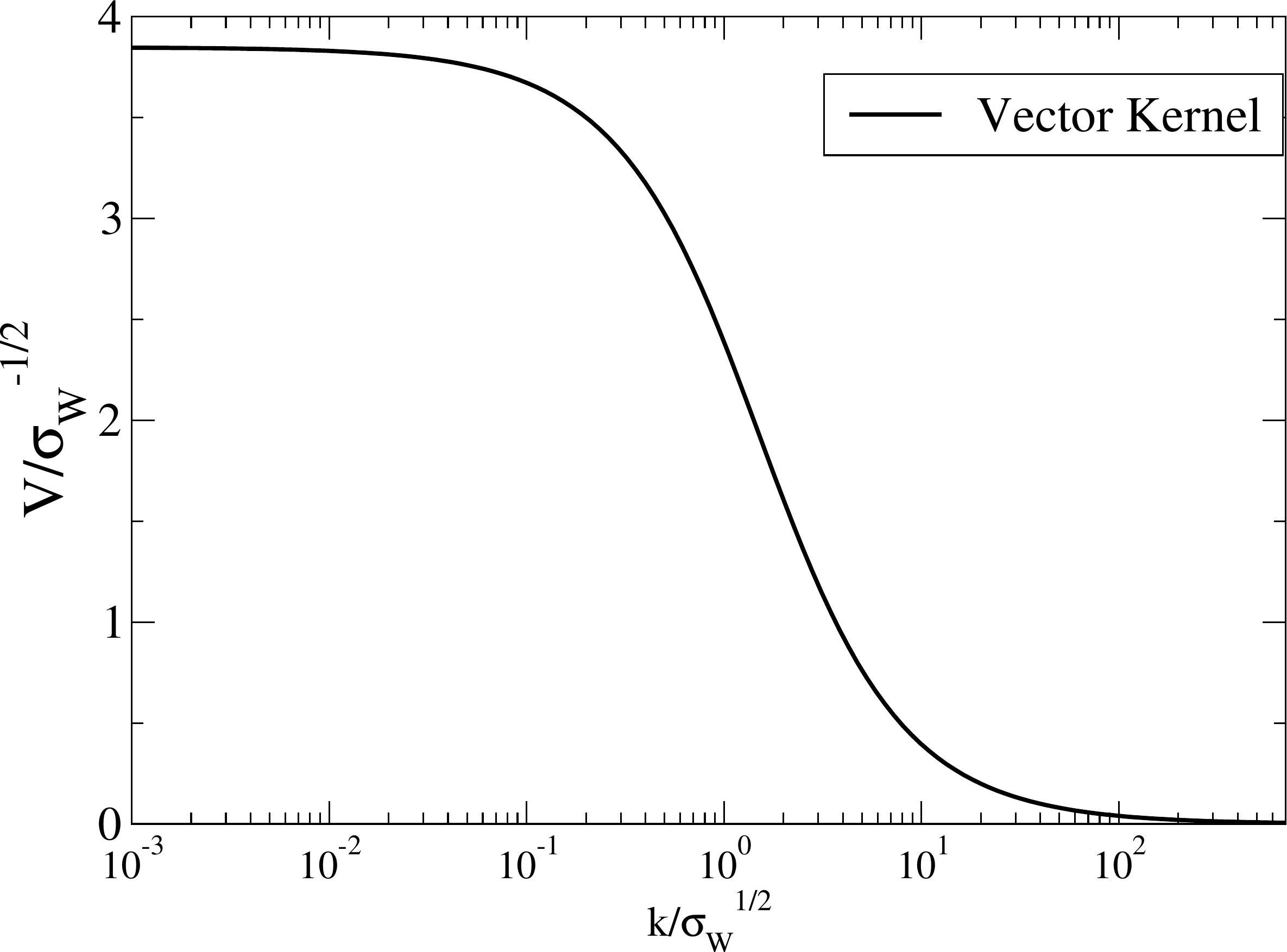}}
\caption[Vector solution function for single-particle gap
equation]{\sl The vector form factor $V(k)$ resulting from the solution of the gap equation
(\ref{v-gap-positive}).}
 \label{fig-V-solution}
\end{figure}

For the trivial solution $S (k) = 0$ of the gap equation (\ref{Gap1}) the equation of motion (\ref{Gap2}) for the vector kernel
reduces to
\begin{align}
\label{v-gap-positive} V(\bk) \, = \, \frac{g}{2} 
\frac{1+C_F V^2(\bk) I(\bk)}{|\bk| +  g C_F V(\bk) I(\bk)} \; ,
 \end{align}
which can be easily solved, yielding
\begin{align}
\label{CH-1195-G27}
V (k) = \frac{k}{g C_F I (k)} \left[ \pm \sqrt{1 + \frac{g^2 C_F I (k)}{k^2}} - 1 \right] \, .
\end{align}
Only the upper sign corresponds to the physical solution since the quark gluon vertex $V (k)$ has to vanish in the limit $g \to 0$. Indeed
for the upper sign we find for small $g$ with the UV-behavior of the gluon loop $I (k)$, Eq.~(\ref{UV-lead}),
\begin{align}
\label{CH-1201-G27A}
V (k) = \frac{g}{2 k} + O \lk \lk \frac{g}{k} \rk^2 \rk \, .
\end{align}
We also observe from the solution (\ref{CH-1195-G27}) that the
limit of small $g$ is equivalent to the limit of large $k$, which
is, of course, a consequence of asymptotic freedom. Thus
Eq.~(\ref{CH-1201-G27A}) gives already the UV-behavior of $V
(k)$. The solution (\ref{CH-1195-G27}) provides also an IR
constant behavior as found in the previous section
\begin{align}
\label{CH-1209-er} V (k = 0) = \frac{1}{\sqrt{C_F I (0)}} =
\text{const.} 
\end{align}
The solution $V (k)$, Eq.~(\ref{CH-1195-G27}), is plotted in
Fig.~\ref{fig-V-solution}. All dimensionful quantities are given
in units of the Wilsonian string tension $\sigma_{\textsc{W}} =
(440 \text{MeV})^2$. In the present calculation the physical
scale is set by the Gribov mass $M_{\textsc{G}}$
(\ref{Gribov-Mass}), which enters the gluon energy
(\ref{Gribov-fit}).
For the Casimir invariant $C_F$ the $SU(3)$ value $C_F=4/3$ is
taken. Moreover, the (running) coupling $g$ (which was
calculated in the Hamiltonian approach in Ref.~\cite{Epple:2006hv}
from the ghost-gluon vertex) is replaced by its infrared value $g \equiv
g(k=0)=8\pi/\sqrt{3 N_{\textsc{C}}}$.

The investigations of the present section show that the coupling
of the quarks to the transverse (spatial) gluons alone is
incapable to induce spontaneous breaking of chiral symmetry. On
the other hand we know from Ref.~\cite{Adler:1984ri} that the
color Coulomb interaction $V_{\textsc{C}}(\bk)$, Eq.~(\ref{Ch-960-22}),
alone does generate spontaneous breaking of chiral symmetry but
not the sufficient amount, as we will see in the next section.

\section{Numerical results}\label{Xnumres}

When the Coulomb potential $V_{\textsc{C}}(\bk)$, Eq.~(\ref{Ch-960-22}), is included the equations of motion (\ref{Gap1}) and (\ref{Gap2})
contain two dimensionful
quantities: the Coulomb string tension $\sigma_{\textsc{C}}$, Eq.~(\ref{394}), and the Gribov mass $M_{\textsc{G}}$, Eq.~(\ref{Gribov-fit}), of the transverse gluon propagator.
As discussed at the end of Sect.~\ref{Yang-Mills-sec} these two quantities are not independent of each other. The
Gribov mass $M_{\textsc{G}}$ can be rather accurately determined on the lattice and we will use its lattice value (\ref{Gribov-Mass}). The Coulomb
string tension is much less accurately determined, see Eq.~(\ref{394}).

\begin{figure}[t]
\centering
\includegraphics[angle=-90,width=.9\linewidth]{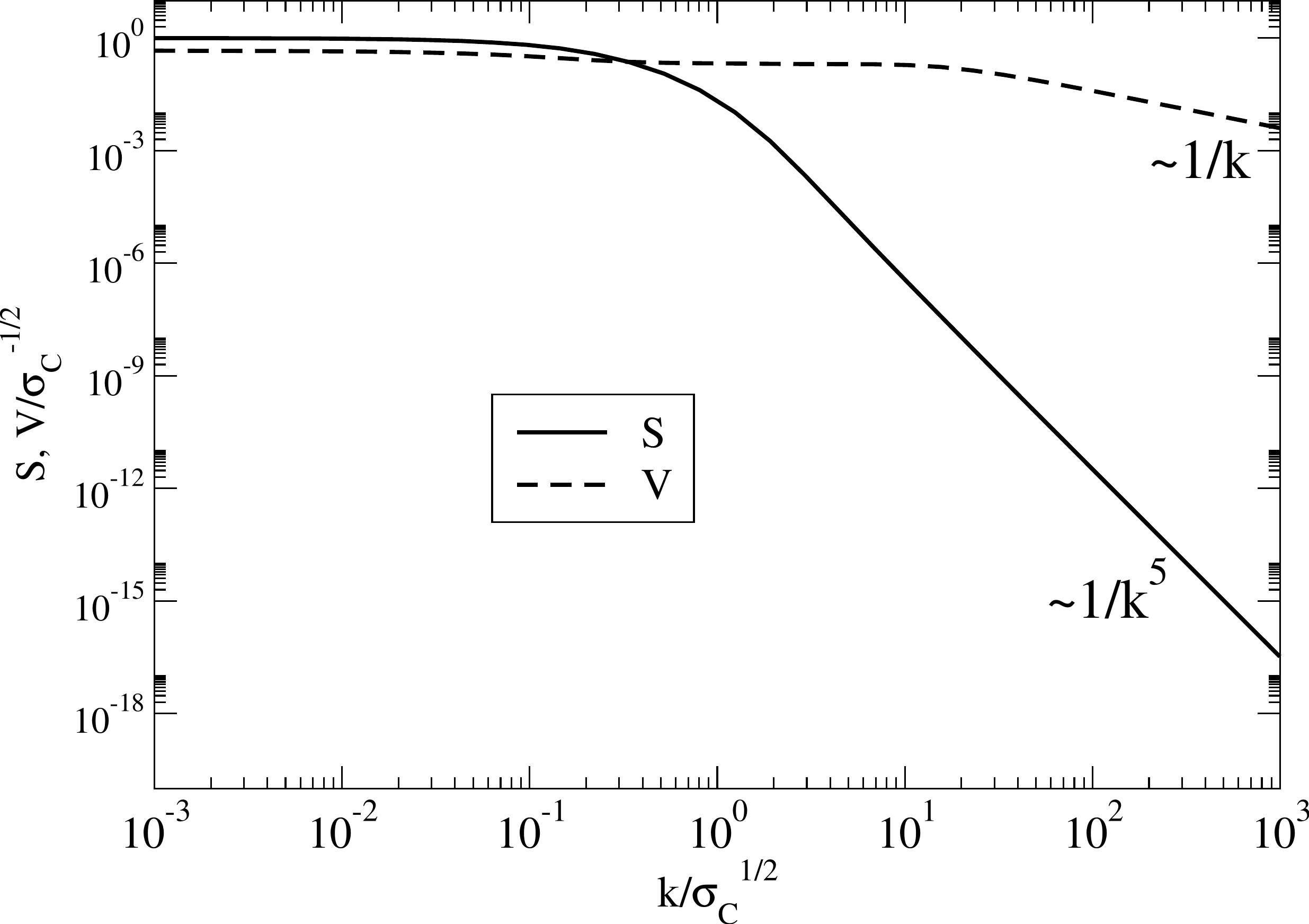}
\caption[Coupled solution with non-perturbative gluon propagator]{\sl Variational kernels $S$ (full curve) and $V$ (dashed curve)
solving the gap equations (\ref{Gap1}), (\ref{Gap2}).}
\label{fig-s-v-asymptotic}
\end{figure}
\begin{figure}[t]
\centering
\includegraphics[width=.9\linewidth]{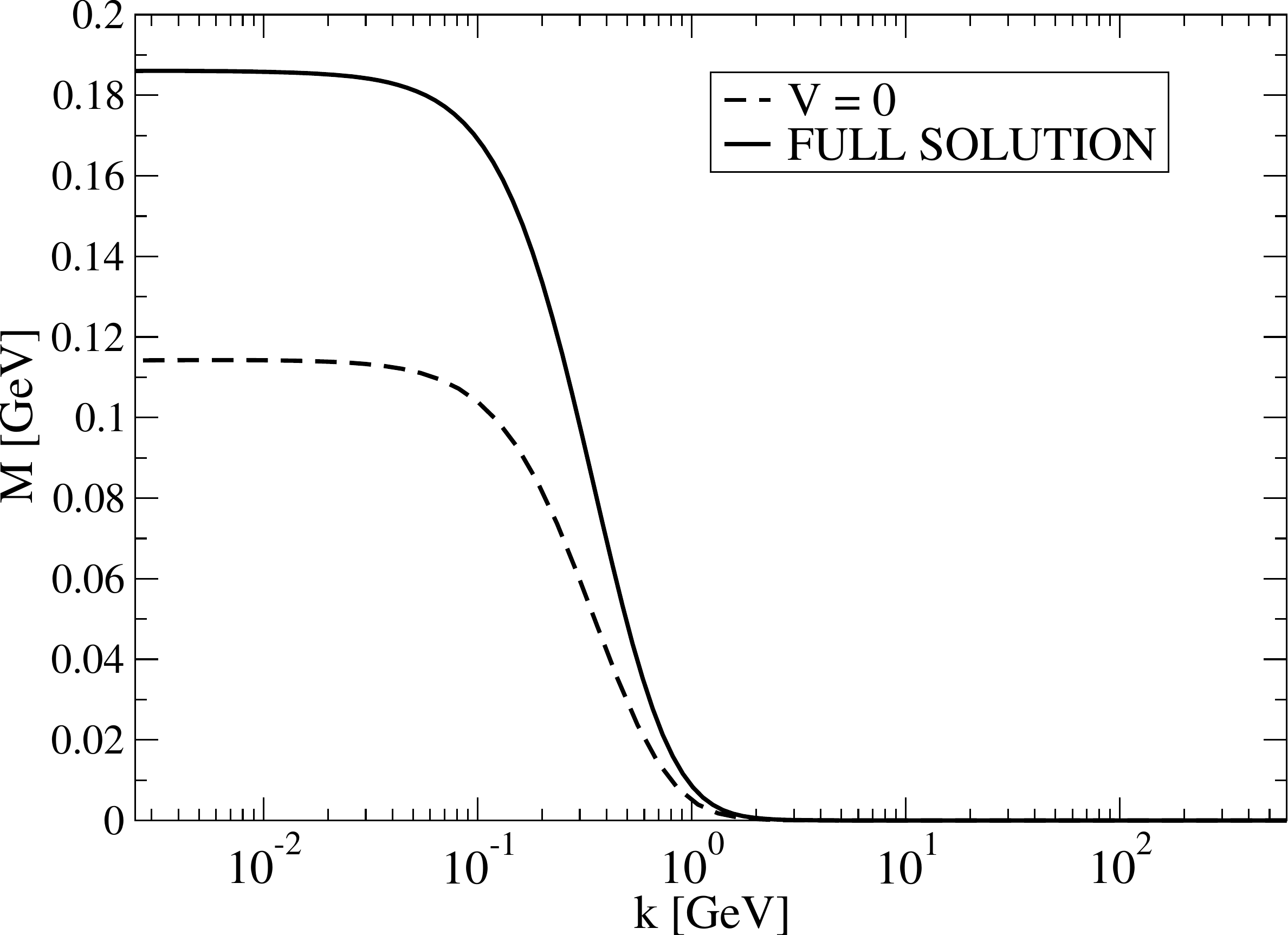}
\caption[Dynamical mass for Adler-Davis and coupled solution]{\sl
The dynamical quark mass $M (\vk)$, Eq.~(\ref{dynam-mass}), for the full solution and for $V=0$ with $\sigma_C = 2 \sigma_W$.} \label{fig-dynam-mass-coupled}
\end{figure}

For the numerical solution of the coupled equations (\ref{Gap1}) and (\ref{Gap2})
all dimensionful quantities are expressed in terms of the
Coulomb string tension
$\sigma_{\textsc{C}}$.
Furthermore, to avoid problems due to the divergence of the Coulomb potential $V_{\textsc{C}} (\vk)$ at $\vk \to 0$ we introduce an IR regulator
$\varepsilon$
\begin{align}
\label{num,-1261}
V_{\textsc{C}} (\vk) \to V_{\textsc{C}} (\vk, \varepsilon) = \frac{8 \pi \sigma_{\textsc{C}}}{\vk^2 \lk \vk^2 + \varepsilon^2 \rk} \, .
\end{align}
Then our solutions $S (\vk)$ and $V (\vk)$ will depend on $\varepsilon$. However, we have tested that the solutions $S (\vk, \varepsilon)$ and
$V (\vk, \varepsilon)$ both converge for $\varepsilon \to 0$. The resulting numerical solutions for $S (k)$ and $V (k)$ are shown in
Fig.~\ref{fig-s-v-asymptotic}. These solutions confirm the asymptotic behavior obtained in Sect.~\ref{Asympt-sect}. From these
solutions one finds the dynamical quark mass $M (k)$, Eq.~(\ref{dynam-mass}), shown in Fig.~\ref{fig-dynam-mass-coupled}. It reaches a plateau
value at small momenta and vanishes for $k \to \infty$, in accordance with asymptotic freedom. The plateau value $M (k = 0)$ defines
the constituent mass, which is obtained as
\begin{align}
\label{num-1271}
M (0) = 132 \, \text{MeV} \sqrt{\sigma_{\textsc{C}} / \sigma_{\textsc{W}}} \, .
\end{align}
For the quark condensate, Eq.~(\ref{chiral-cond}), we find
\begin{align}
\label{num-1276}
\langle \overline{\psi} \psi \rangle \simeq \lk - 135 \, \text{MeV} \sqrt{\sigma_{\textsc{C}} / \sigma_{\textsc{W}}}  \rk^3 \, .
\end{align}
Using $\sigma_{\textsc{C}} = (2 \ldots 3) \sigma_{\textsc{W}}$ we obtain
\begin{align}
\label{num-1281}
M (0) & \simeq (186 \ldots 230) \, \text{MeV} \nonumber\\
\langle \bar{\psi} \psi \rangle & \simeq - (191 \ldots 234 \, \text{MeV})^3 \, .
\end{align}
While the obtained mass is somewhat smaller than the constituent mass of the light quark flavors, which is about 300 MeV,
the obtained chiral
condensate compares more favorable with the phenomenological value of $\langle \bar \psi \psi \rangle = (- 230 \, \text{MeV})^3$.

\begin{figure}[t]
\centering
\includegraphics[angle=-90,width=.9\linewidth]{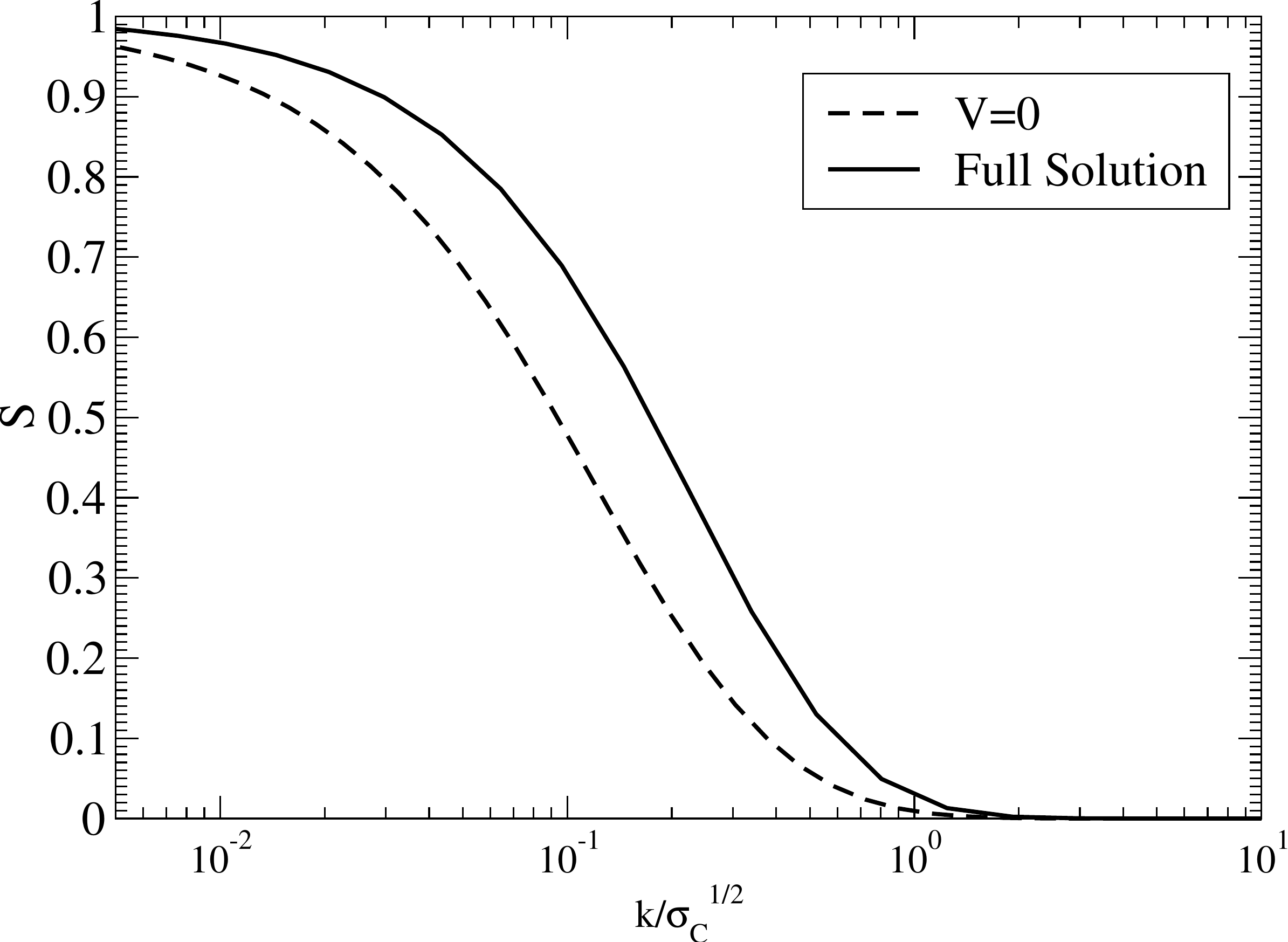}
\caption[Comparison Adler-Davis and coupled solution with
non-perturbative gluon propagator]{\sl Variational kernel $S$
comparing the solutions with $V=0$ and $V\neq 0$.}
\label{fig-s-comparison}
\end{figure}

Let us now compare our results with those obtained when the coupling of the quarks to the transverse gluons is neglected, $V (k) = 0$, i.e. when the BCS-type quark wave
functional is used, Ref.~\cite{Adler:1984ri}. Figs.~\ref{fig-dynam-mass-coupled} and \ref{fig-s-comparison} show the dynamical mass $M (\vk)$ and the 
scalar form factor $S (\vk)$, respectively, for both cases. When the coupling to the transverse gluons is neglected
both $M(\vk)$ and $S(\vk)$ are substantially reduced. One finds then
\begin{align}
\label{num-1292}
M (0) & \simeq 84 \, \text{MeV} \sqrt{\sigma_{\textsc{C}} / \sigma_{\textsc{W}}}  \nonumber\\
\langle \overline{\psi} \psi \rangle & \simeq - \lk 113 \, \text{MeV} \sqrt{\sigma_{\textsc{C}} / \sigma_{\textsc{W}}} \rk^3 \, .
\end{align}
Compared to these results the inclusion of the coupling of the quarks to the transverse gluons, i.e. of the vector kernel $V (k)$,
increases the quark condensate $\langle \overline{\psi} \psi \rangle $, Eq.~(\ref{num-1276}),
by 20$ \%$  and the constituent mass, Eq.~(\ref{num-1271}), by
60 $ \%$ .

\begin{figure}[t]
\centering
\includegraphics[width=.9\linewidth]{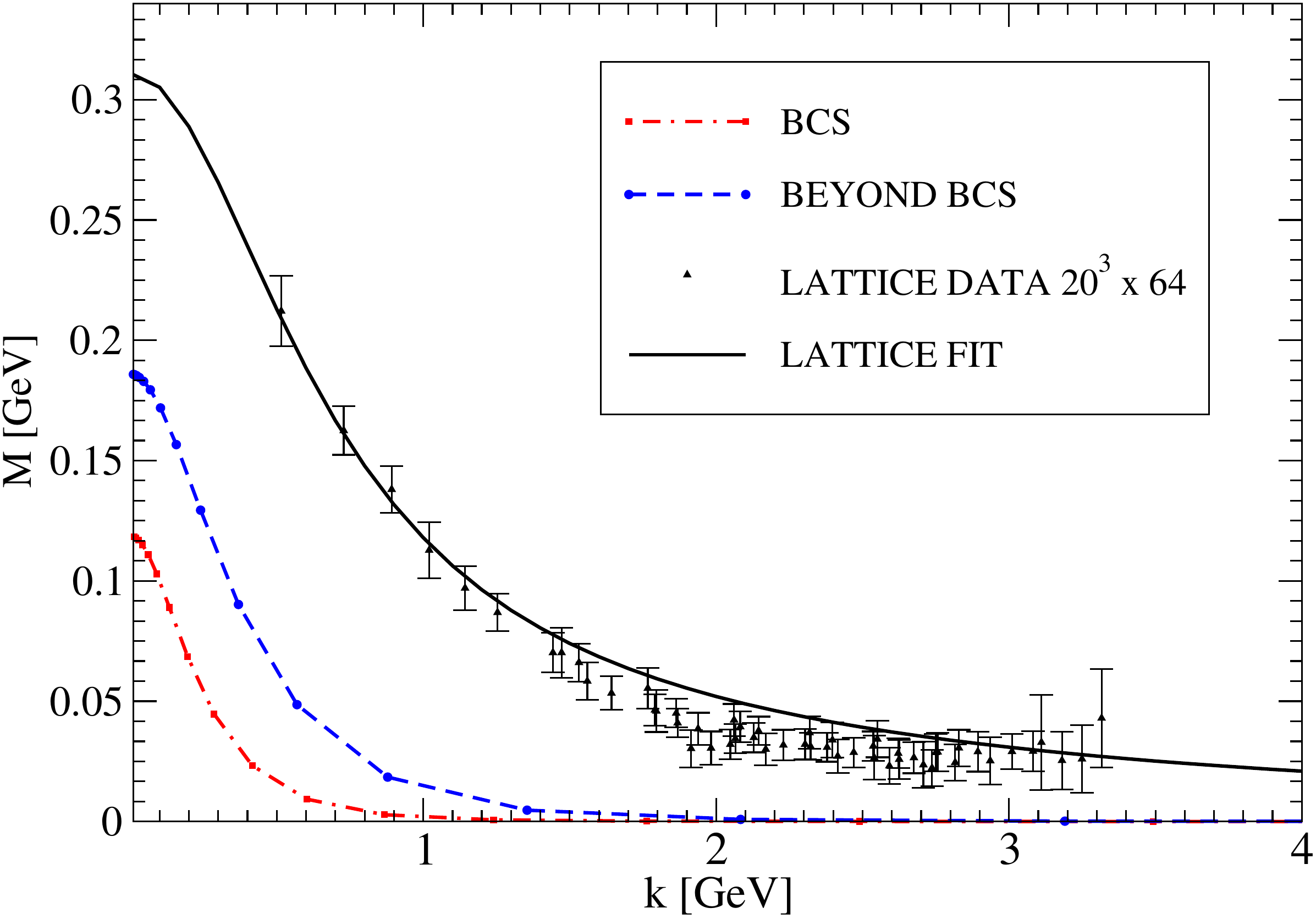}
\caption[Dynamical mass compared to lattice]{\sl
The dynamical quark mass $M (k)$, Eq.~(\ref{dynam-mass}), for $\sigma_{\textsc{C}} = 2 \sigma_{\textsc{W}}$ and $\sqrt{\sigma_{\textsc{W}}} = 440 $ MeV
compared to the lattice data obtained in Ref.~\cite{Burgio:2012ph}.  The dashed curve is obtained with the quark-gluon coupling
included in the wave functional while the dotted curve is obtained with the BCS wave functional.} \label{fig-dynam-mass-lattice}
\end{figure}

In Fig.~\ref{fig-dynam-mass-lattice} we compare our results for
the dynamical mass to the lattice data obtained recently in Ref.~\cite{Burgio:2012ph}. As one observes the shape of
the momentum dependence is reproduced but the absolute values are
still too small.

\section{Unquenching the Gluon Propagator}\label{XIunqu}

So far all calculations were done in the quenched approximation, i.e. the gluon propagator and the Coulomb kernel were taken from
the pure Yang-Mills sector and used as input for the treatment of the quark sector. In a fully unquenched calculation the variation
would be carried out at
the same time with respect to all variational kernels and the resulting equations of motion had to be solved self-consistently. Below we give an estimate of the
unquenching effects by calculating the corrections to the gluon propagator but using in these corrections the
quenched gluon propagator as input.

The unquenching arises from two sources: First, from those fermionic
contributions to the energy density, which depend on the gluon propagator, see Eqs.~(\ref{qgc-expect}) and (\ref{Coulomb-energy}).
Second from the norm of the fermionic wave functional (the fermion determinant), see Eq.~(\ref{557}), which does depend
on the gauge field. In a fully unquenched calculation the fermionic wave functional must not be normalized separately from the gluonic one, as
we did in the present calculation given above. We will investigate both effects separately.
\vspace{0.5cm}
\subsection{Quark energy contributions}

In Ref.~\cite{Heffner:2012sx} it was shown that the gluonic Coulomb term is irrelevant for the Yang-Mills sector.
We expect that this is also true for the contribution
of the quark Coulomb energy (\ref{Coulomb-energy}) to the unquenching of the gluon propagator. This is because the
Coulomb potential (\ref{384}) depends
only implicitly on the gluon propagator and variation of Eq.~(\ref{Coulomb-energy}) with respect to the
gluon propagator gives rise to more than two loops. We are then left with the quark energy $\langle H_F \rangle$, Eq.~(\ref{qgc-expect}).
Variation of this quantity with respect to the gluon propagator $\omega^{- 1}(\bk)$ yields
\begin{widetext}
\begin{align}
 &\frac{\delta}{\delta \omega^{-1}(\bk)}\left[\frac{\langle H_F \rangle}{\delta^{3}(0)} \right] \equiv \Delta
(\bk) = N_{\textsc{C}} C_F \int \dbar^3 p \frac{1}{1+S^{\ast}(\bp) S(\bp)+ R(\bp)} \times \nonumber\\
&\times \left[ g \, \left(V^{\ast}(\bp,\boldsymbol{\ell}) +
V(\bp,\boldsymbol{\ell})\right) + 2 \frac{V(\bp,\boldsymbol{\ell}) V^{\ast}(\boldsymbol{\ell},\bp)}{1+S^{\ast}(\bp) S(\bp)+ R(\bp)} \right] \left[1+ (\hat{\bp} \cdot \hat{\bk})(\hat{\boldsymbol{\ell}} \cdot
\hat{\bk}) \right] \; ,
\label{Int-Unquench}
\end{align}
\end{widetext}
with $\boldsymbol{\ell} = \bp-\bk$.
This gives an extra contribution to the gluonic gap equation, Ref.~\cite{Feuchter:2004mk}, which then reads
\begin{align}
\label{ch-1366-32}
\omega^2 (\bk) = \bk^2 + \chi^2 (\bk) + \Delta (\bk) = \omega^2_{\textsc{YM}}(\bk) + \Delta(\bk) \, ,
\end{align}
where $\omega_{\textsc{YM}}(\bk)$ is the gluon energy in the pure Yang-Mills case (in the previous section this quantity was
called $\omega(\bk)$).
Obviously, the unquenching correction $\Delta (\bk)$, Eq.~(\ref{Int-Unquench}), disappears when the coupling of the quarks
to the transverse gluons is neglected, $V (\vp, \boldsymbol{\ell}) = 0$.

The quark contribution $\Delta (\bk)$, Eq.~(\ref{Int-Unquench}), is UV-divergent. It is straightforward to extract its
divergence structure
\begin{align}
\label{ch-1375-32}
\Delta_{\textsc{div}} (k, \Lambda) = \sim \Lambda^2 + \sim \Lambda + \sim k^2 \ln \Lambda \, ,
\end{align}
where $\Lambda$ is the 3-momentum cut-off. The quadratic and linear divergence disappear when the gap equation is
renormalized by subtracting it at a renormalization scale. More elegantly these terms as well as the logarithmic divergence
are eliminated by adding appropriate counter terms to $H_{\textsc{F}}$, Eq.~(\ref{167-x4}), analogous to the renormalization in the gluon sector,
see Refs.~\cite{Reinhardt:2007wh}, \cite{Epple:2007ut}. Here we will just consider the finite part of $\Delta (\bk)$, Eq.~(\ref{Int-Unquench}), and use
the renormalization condition
\begin{align}
\label{ch-1384-32}
\frac{\omega^2(\zeta, \Lambda)}{\zeta^2} = \frac{\omega^2_{\textsc{YM}}(\zeta)}{\zeta^2} \, ,
\end{align}
and fix the renormalization point $\zeta$ in the ultraviolet, which is justified since for large momenta the quark
vacuum becomes bare.

\begin{figure}[t]
\centering
\includegraphics[angle=0,width=.7\linewidth]{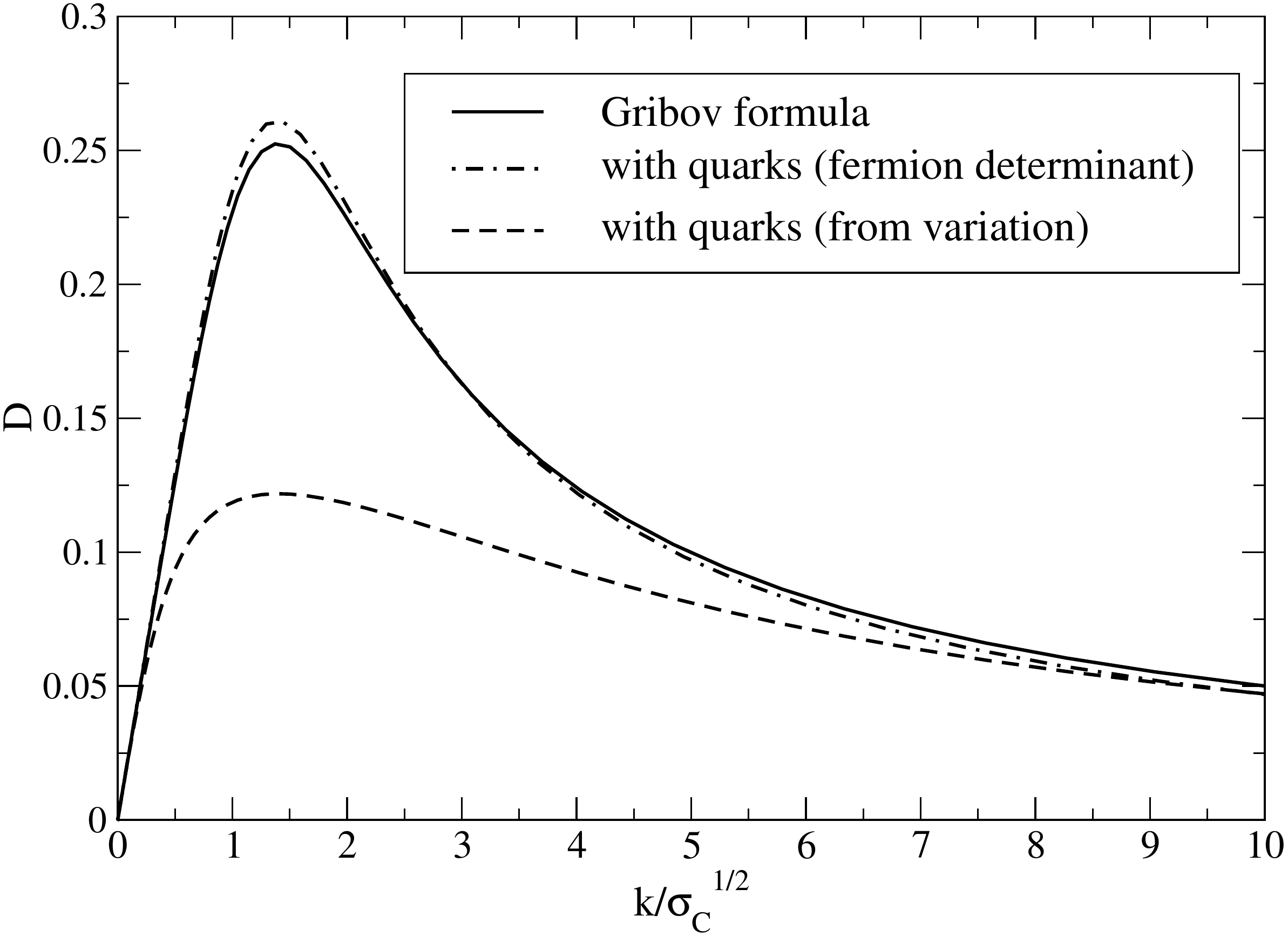}
\caption[Unquenching effects in the gluon propagator.]{\sl The full curve is the Gribov form. The dash-dotted curve is the unquenching effect considered in Eq.~(\ref{ch-1366-32}).
The dashed line shows the result (\ref{Sigma-Res}).}
\label{fig-Unquenching}
\end{figure}

To estimate the unquenching effect due to the quark energy contribution
 we assume for $\omega_{\textsc{YM}} (k)$ the Gribov formula
(\ref{Gribov-fit}) with the same Gribov mass $M_{\textsc{G}} = 880$ MeV as used above.
The resulting gluon propagator $D(k) = 1 / (2 \omega (k))$ is shown in Fig.~\ref{fig-Unquenching} together with the
quenched result. It is seen that the unquenching decreases the gluon propagator in the mid-momentum regime but leaves the UV
and IR asymptotic behaviour unchanged. Unfortunately, it is the mid-momentum regime which is relevant for the hadron physics and also for the
deconfinement phase transition. Therefore unquenching seems to
 be important for a realistic description of hadrons and
the deconfinement transition.

\subsection{The Fermion Determinant}

As already discussed above, in a fully unquenched calculation only the total wave functional of QCD can be normalized
while the norm of the quark wave functional becomes a dynamical object due to its dependence on the gauge field. The norm
of our quark wave functional (\ref{coordinate-wave}) is given by Eq.~(\ref{557})
\begin{align}
\label{ch-1423-33}
\langle \phi_{\textsc{F}} | \phi_{\textsc{F}} \rangle = \mbox{Det} \Omega = \exp (\mbox{Tr} \ln \Omega) \, ,
\end{align}
where the matrix $\Omega$ is defined by Eq.~(\ref{549}) and is a 
functional of the gauge field $\vA$.
By Wick's theorem this quantity arises as factor in all fermionic
expectation values.
In the scalar product of the QCD
wave functionals (\ref{280-y2}) $\mbox{Det} \Omega$ can be considered as part of the Yang-Mills wave functional
$| \phi_{\textsc{YM}} (\vA) |^2$. To keep the gluonic functional integral Gaussian we expand $\mbox{Tr}\ln \Omega$
up to second order in the gauge field. With (see Eqs.~(\ref{485}) and (\ref{549}))
\begin{align}
\label{ch-1438-2}
\Omega = \Omega_0 + \Omega_1 \cdot A
\end{align}
this yields
\begin{widetext}
\begin{align}
\label{ch-1443-ac}
\mbox{Tr} \ln \Omega &= \mbox{Tr} \ln \Omega_0 + \mbox{Tr} \lk \Omega^{- 1}_0 \Omega \cdot A \rk - \frac{1}{2} \mbox{Tr}
\lk \Omega^{- 1}_0 \Omega_1 \cdot A \, \Omega^{- 1}_0
\Omega_1 \cdot A \rk 
= \text{const.} + \frac{1}{2} \int A \, \Sigma \, A \, .
\end{align}
\end{widetext}
The zeroth order term is an irrelevant constant which can be absorbed into the 
overall normalization of the QCD wave functional.
The linear term vanishes due to the color trace and the quadratic term contributes directly to the gluon energy. Due to the
transversality of the gauge field and the absence of external color fields we have
\begin{align}
\label{ch-1452-ef}
\Sigma^{ab}_{ij} (\vk) = \delta^{ab} t_{ij} (\bk) \Sigma (\bk) \, .
\end{align}
The explicit calculation yields
\begin{widetext}
\begin{align}
\label{Xi-Res}\Sigma (\bk) &= - \frac{1}{2} \int \dbar^3 p \,
\frac{V^2(\bp,\bp+\bk)}{(1+S^2(\bp+\bk))(1+S^2(\bp))}
\left(1+S(\bp) S(\bp+\bk)\right) 
 (1+(\hat{\bk}\cdot \hat{\bp}) (\hat{\bk}\cdot
(\widehat{\bp+\bk}))) \; . 
\end{align}
\end{widetext}
Taking into account the unquenching effects from the quark loop the gluon propagator is then given by
\begin{align}
\label{Sigma-Res}
{\omega} (\bk) = \omega_{\textsc{YM}}(\bk) + \Sigma (\bk) \; .
\end{align}
Again the unquenching disappears, $\Sigma (\vk) = 0$, when the coupling of the quarks to the spatial gluons is
switched off in the wave functional, $V (\bk) = 0$. The integral $\Sigma (\vk)$, Eq.~(\ref{Xi-Res}), is linearly divergent.
Carrying out the renormalization as in the previous subsection we find for the (partially) quenched gluon propagator the
result shown in fig.~\ref{fig-Unquenching}.
Again, the gluon propagator is affected by the quarks only in the intermediate momentum region.
However, now the propagator is increased, although the amount of increase is much less than the decrease found in the previous subsection from the quark-gluon
coupling energy $\langle H_F \rangle$, Eq.~(\ref{Int-Unquench}).  From this we can conclude that the unquenching reduces the gluon propagator in the
mid momentum regime.
In a self-consistent solution both of these unquenching effects combine non-trivially in the
gap equations. Due to the net reduction of the gluon propagator by the unquenching, we expect that in a fully self-consistent
calculation the unquenching effects are less dramatic than found above but may still be essential.

\section{Summary and Conclusions}\label{XIIsum}

The variational approach to Yang-Mills theory in Coulomb gauge developed previously in Ref.~\cite{Feuchter:2004mk} has been
extended to full QCD. The QCD Schr\"odinger equation has been variationally solved in the quenched approximation using an ansatz
for the quark wave functional, which explicitly includes the coupling of the quarks to the spatial gluons and thus goes beyond
previously used BCS type quark wave functionals. For the Yang-Mills sector we have used the vacuum wave functional determined previously in
\cite{Feuchter:2004mk} and \cite{Epple:2006hv} as input.

Our quark wave functional contains two variational kernels: One scalar kernel $S (\vk)$, which is related to the quark condensate
and occurs already in the BCS type wave functionals, and a vector kernel $V (\vk)$, which represents the form factor of the
quark gluon coupling. The equations of motion following from the variational principle for these kernels have been solved analytically in the
infrared and in the ultraviolet and numerically in the whole momentum regime. Both kernels are infrared finite and vanish at large momenta
in accordance with asymptotic freedom. We have shown that neglecting the color Coulomb potential the coupling of the quarks
to the spatial
gluons is not capable of triggering spontaneous breaking of chiral symmetry and produces always a vanishing scalar kernel $S (\vk) = 0$.
When the confining color Coulomb potential is included the coupling of the quarks to the gluons substantially enhances the amount
of chiral symmetry breaking towards the phenomenological findings. The quark condensate is increased by about 20 $\%$ and compares
favourably with the phenomenological values. Although the constituent quark mass is increased by about 60 $\%$ due to the coupling of the
quarks to the spatial gluons, the value found is still somewhat small.

One may speculate where the missing chiral strength is lost in the
present approach. Certainly we have used a couple of approximations but given the success of the present approach in the pure Yang-Mills
sector one would perhaps expect a better agreement with the phenomenological data. First one should remark that the lattice
calculations done in Coulomb gauge, Ref.~\cite{Burgio:2012ph}, show that the running quark mass $M (\vk)$ is smaller in the quenched calculation
compared to the dynamical one. But this effect is only of the order of a few percent. Next one may question the additional approximation
$1 / (\Id + K \overline{K}) \to 1 / (\Id + \langle K \overline{K} \rangle_{\textsc{G}})$ we have used in the quark sector when calculating the gluonic
expection values of fermionic operators. We do not expect that this approximation makes big quantitative changes, since this
replacement is exact for the scalar kernel $K_0 \sim S (\vk)$, which dominates the chiral properties. One may then ponder on our
ansatz for the quark wave functional, Eq.~(\ref{coordinate-wave}).
This wave functional, being given by an exponent which is bilinear in the
quark field, represents the most general Slater determinant. One certainly does not want to abandon the determinantal states in order not
to lose Wick's theorem. However, we have assumed for the kernel $K$ in the exponent of the quark wave functional an expansion in
powers of the gauge field and restricted this expansion in linear order. This is certainly a rather crude
 approximation. It is
straightforward to extend the present approach by keeping in the kernel $K$ (\ref{485}) higher powers of the spatial gluon field. Then the
fermionic part of the present calculation does not change at all. What changes is the gluonic expectation value, which, however,
can be done by using Wick's theorem since the employed gluonic wave functional is Gaussian. We expect a substantial improvement by
including terms in the kernel $K$, which are second order in the gauge field. This will introduce further variational kernels, which
can only improve the results towards the exact ones.

In principle, one can go beyond the determinantal quark wave functional (\ref{coordinate-wave}) and include in its exponent e.g. four fermion
operators. Such a wave functional can be handled by means of the approach based on Dyson-Schwinger equation techniques and developed
in Ref.~\cite{Campagnari:2010wc} for the treatment of non-Gaussian wave functionals 
in the Yang-Mills sector. An extension of this approach to full QCD is in progress \cite{R3}.

Finally in the last part of our paper we have considered a partial unquenching by including the change of the gluonic energy due to the
presence of the quarks. Our results show that unquenching reduces the static gluon propagator in the mid-momentum regime.
The unquenching effects disappear when the coupling of the quarks to the spatial gluons is neglected.

Some results of the present paper were used in Ref.~\cite{Pak:2013cpa} to study the influence of spatial
gluons on the chiral symmetry patterns of the high-spin meson spectrum. 

\begin{acknowledgements}
Discussions with G.~Burgio, D.~Campagnari, M.~Quandt and P.~Watson
are greatly acknowledged. This work was supported by BMBF 06TU7199, 
by the Europ\"aisches Graduiertenkolleg ``Hadronen im Vakuum,
Kernen und Sternen'', and by the Graduiertenkolleg
``Kepler-Kolleg: Particles, Fields and Messengers of the
Universe''. M.~P. acknowledges support 
by the Austrian Science Fund (FWF)
through the grant P21970-N16.
\end{acknowledgements}

\bibliographystyle{utphys}
\bibliography{Biblio2}

\end{document}